%% file: draft.tex
\newcommand{\final}{0}

\documentclass[final,3p,times,sort&compress]{elsarticle}

\input{tex/package}
\input{tex/macros}

\journal{Computer Physics Communications: https://doi.org/10.1016/j.cpc.2024.109187}

\begin{document}

\begin{frontmatter}

  \title{{XLB}: {A} differentiable massively parallel lattice {B}oltzmann library in {P}ython}
\affiliation[a]{organization={Autodesk Research},
            addressline={661 University Avenue},
            city={Toronto},
            postcode={M5G 1M1}, 
            state={ON},
            country={Canada}}
\author[a]{Mohammadmehdi Ataei\corref{author}}
\author[a]{Hesam Salehipour}

\cortext[author] {Corresponding author.\\\textit{E-mail address:} mehdi.ataei@autodesk.com}

\input{tex/abstract}


\end{frontmatter}






\input{tex/intro}
\input{tex/lbm}
\input{tex/impl}
\input{tex/benchmarks}

\input{tex/performance}
\input{tex/sciML.tex}
\input{tex/conclusion}
\input{tex/ack}


\bibliographystyle{elsarticle-num}
\bibliography{ref}


\end{document}

%% file: tex/package.tex
\usepackage{booktabs} 
\usepackage[ruled,linesnumbered]{algorithm2e}
\usepackage{float}
\usepackage{subcaption} 
\usepackage{xcolor}
\usepackage{dirtree}                    
\usepackage{amsmath}
\usepackage{mathrsfs}
\usepackage{bm}
\usepackage{tikz}
\usepackage{subcaption}
\usepackage{tikz-3dplot}
\usepackage{tabularx}
\usepackage{hyperref}
\hypersetup{
    colorlinks=true,
    linkcolor=blue,
    filecolor=magenta,      
    urlcolor=cyan,
    pdfpagemode=FullScreen,
    }

\definecolor{codegreen}{rgb}{0,0.6,0}
\definecolor{codegray}{rgb}{0.5,0.5,5}
\definecolor{codepurple}{rgb}{0.58,0,0.82}
\definecolor{backcolor}{rgb}{0.95,0.95,0.92}

\usetikzlibrary{arrows.meta}                 
\usepackage{amssymb}



%% file: tex/macros.tex
\definecolor{mehdiColor}{rgb}{1,0.43,0.43}
\newcommand{\mehdi}[1]{{\color{mehdiColor} Mehdi: #1}}

\definecolor{hesamColor}{rgb}{0.0, 0.7, 0.7}
\newcommand{\HS}[1]{{\color{hesamColor} #1}}

\newcommand{\warning}[1]{{\emph{\color{red} #1}}}

\newcommand{\nothing}[1]{}

\ifthenelse{\equal{\final}{1}}
{
\renewcommand{\mehdi}[1]{}
\renewcommand{\HS}[1]{}

\renewcommand{\warning}[1]{}

}
{}

%% file: tex/abstract.tex
\begin{abstract}
The lattice Boltzmann method (LBM) has emerged as a prominent technique for solving fluid dynamics problems due to its algorithmic potential for computational scalability. We introduce XLB library, a Python-based differentiable LBM library based on the JAX platform. The architecture of XLB is predicated upon ensuring accessibility, extensibility, and computational performance, enabling scaling effectively across CPU, TPU, multi-GPU, and distributed multi-GPU or TPU systems. The library can be readily augmented with novel boundary conditions, collision models, or multi-physics simulation capabilities. XLB's differentiability and data structure is compatible with the extensive JAX-based machine learning ecosystem, enabling it to address physics-based machine learning, optimization, and inverse problems. XLB has been successfully scaled to handle simulations with billions of cells, achieving giga-scale lattice updates per second. XLB is released under the permissive Apache-2.0 license and is available on GitHub at https://github.com/Autodesk/XLB.
\end{abstract}

%% file: tex/intro.tex
\section{Introduction}
\label{sec:intro}

In recent years, domain-specific libraries that are built on top of compiler technologies such as XLA and MLIR \cite{zhang2023compiler,githubGitHubOpenxlaxla,lattner2020mlir,lattner2021mlir} have gained substantial attention. These libraries mostly offer high-level programming interfaces while ensuring efficient execution by targeting specialized hardware like GPUs and TPUs. Libraries such as JAX~\cite{jax2018github}, PyTorch~\cite{NEURIPS2019_9015}, Triton~\cite{openaiIntroducingTriton}, and TensorFlow~\cite{tensorflow2015-whitepaper} exemplify this approach. This trend is largely fueled by the increasing interest in machine learning and its various applications. Although predominantly employed in machine learning tasks, these libraries are also valuable for scientific computing and physics-based machine learning applications. They facilitate high-performance computation in Python, a language widely favored for its readability and ease of use, and provide tools for performing automatic differentiation, allowing for the application of machine learning within scientific domains. Numerous specialized libraries, such as JAX-Fluids~\cite{bezgin2023jax}, JAX-CFD~\cite{kochkov2021machine} (see also \cite{wang2022tensorflow}), BRAX~\cite{brax2021github}, JAX-MD~\cite{schoenholz2020jax}, JAX-FEM~\cite{xue2023jax}, Phiflow~\cite{holl2020phiflow}, TaichiDiff~\cite{hu2019difftaichi} and Taichi-LBM3D~\cite{yang2022taichi}, have been developed on top of these platforms.

Recently, machine learning techniques have become increasingly recognized as valuable tool in scientific domains, ranging from molecular interactions and protein synthesis to astronomical and geophysical phenomena~\cite{fluke2020surveying,baron2019machine,wang2018deepmd,zhang2020dp,jumper2021highly, wang2022deep, salehipour2019deep}. Machine learning algorithms have been applied to many disciplines in physical sciences~\cite{carleo2019machine,brunton2021applying}, to interpret large data sets, or uncover new scientific knowledge from raw data~\cite{jia2021physics,radovic2018machine,vasudevan2021machine,hochreiter2018machine}. When combined with traditional scientific computing methods, machine learning can significantly enhance both the speed and accuracy of simulations and analyses~\cite{edelen2020machine,chen2022wavey}. In fluid dynamics, the application of machine learning techniques has led to novel advancements in improving turbulence models, unsteady flow prediction, optimizing flow configurations, or predicting complex flow phenomena with high accuracy~\cite{vinuesa2022enhancing,halder2024reduced,kochkov2021machine,fukami2020assessment, salehipour2019deep}. Differentiable fluid simulations when combined with deep learning approaches have demonstrated progress in a range of applications, from fast and accurate fluid flow prediction to learned turbulence models, shape optimization, and fluid control~\cite{schenck2018spnets,takahashi2021differentiable,wandel2020learning,belbute2020combining,brahmachary2023unsteady,holzschuh2023score,um2020solver,list2022learned,chen2021numerical}. 

The Lattice Boltzmann Method (LBM), originating from the kinetic gas theory, has emerged in recent decades as a widely accepted technique for tackling complex fluid dynamics problems. Its algorithmic foundation, built upon a simple yet effective `collide-and-stream' mechanism on Cartesian grids, makes it highly parallelizable and exceptionally suitable for use with GPUs and TPUs. LBM has proven to be a promising methodology in simulating laminar and turbulent flows, subsonic and supersonic flows, flow through porous media, free-surface and multiphase flows, as well as complex flows involving fluid-solid interaction, combustion, and foaming phenomena~\cite{huang2015multiphase,renard2021improved,liu2016multiphase,grunau1993lattice,mccracken2005multiple,ataei2021lbfoam,ataei2019numerical,thurey2004free,ginzburg2003lattice,ataei2022hybrid,schreiber2011free, chiavazzo2009combustion}. Recently, the integration of machine learning methods with LBM-based simulations has opened new frontiers in the field, providing enhanced capabilities such as the development of fast surrogate models and improved acoustic predictions~\cite{ruttgers2020prediction, chen2021compressed, golsanami2022characterizing, zhu2021numerical, da2020ml, hennigh2017lat, bedrunka2021lettuce, guo2016convolutional}.

In this paper, we introduce XLB, a differentiable massively parallel LBM library based on JAX, aimed to be used for large-scale fluid simulations, optimization, and physics-based machine learning. XLB has been developed with three key objectives. First, to prioritize accessibility, XLB employs Python and offers an interface that closely resembles Numpy, thereby ensuring ease-of-use and enabling quick adoption by a broad user base. Second, to emphasize extensibility, XLB adopts an object-oriented programming model, which allows users to effortlessly augment the library's capabilities and tailor it to diverse research needs. Lastly, despite its user-friendly design, XLB does not sacrifice performance; it is engineered for high performance and scalability, making it suitable for both entry-level usage and advanced, resource-intensive applications.

While open-source LBM libraries such as Palabos~\cite{latt2021palabos} and OpenLB~\cite{krause2021openlb} as well as high-performance programming models suitable for LBM calculations (see e.g.~\cite{meneghin2022neon}) exist, the majority are written in low-level programming languages like C/C++. These libraries have high learning curves that make them less than ideal for rapid research and prototyping where the application itself may demand more focused investigation (rather than solver configuration and details). Hence, integration of these tools with available machine learning libraries such as PyTorch or JAX becomes complicated rendering the development of a unified computational model for physics-based machine learning very challenging. XLB has been designed to address these issues by providing an accessible and scalable solution for such applications. It successfully mitigates the performance shortfalls commonly associated with Python-based solutions, primarily through just-in-time (JIT) compilation and distributed computing, positioning it as a compelling choice for research labs requiring a blend of high-performance and ease-of-use. Additionally, XLB's adoption of the permissive Apache License, in contrast to the more restrictive GPL-based licenses of other LBM libraries, enhances its appeal for broader use in both academic and commercial applications.

The most closely related software to XLB is Lettuce~\cite{bedrunka2021lettuce}, a PyTorch-based library that integrates LBM simulations with the PyTorch deep learning ecosystem. Nevertheless, XLB provides superior performance and offers the added advantage of being scalable on distributed multi-GPU architectures. Furthermore, XLB's intuitive and flexible programming model, derived from JAX's Numpy-like interface, affords users a more straightforward path for extending the library's functionality. 

The remainder of this paper is organized as follows. In Section~\ref{sec:lbm}, we provide some basic preliminaries on LBM. Section~\ref{sec:impl} outlines the programming model of XLB. Benchmark results to validate and verify XLB functionalities are presented in Section~\ref{sec:benchmarks}. Performance evaluations on single-GPU, multi-GPU, and distributed multi-GPU systems are discussed in Section~\ref{sec:performance}. Finally, Section~\ref{sec:sci_ml} showcases examples of using XLB for physics-based machine learning applications. Our concluding remarks are summarized in Section~\ref{sec:conclusion}.

%% file: tex/lbm.tex
\section{Lattice Boltzmann method}
\label{sec:lbm}

\begin{figure}[H]
    \centering
    \begin{subfigure}{0.3\textwidth}
        \begin{tikzpicture}[>=Stealth]
            \draw[blue!20, opacity=0.5] (-1,-1) -- (-1,1) -- (1,1) -- (1,-1) -- cycle;
            \draw[green!20, opacity=0.5] (0,-1) -- (0,1);
            \draw[red!20, opacity=0.5] (-1,0) -- (1,0);
            \draw (-1,-1) rectangle (1,1);
            \foreach \x/\y in {1/0, 0/1, -1/0, 0/-1, 1/1, 1/-1, -1/1, -1/-1} {
              \draw[->, thick] (0,0) -- (\x,\y);
            }
        \end{tikzpicture}
    \end{subfigure}
    \begin{subfigure}{0.3\textwidth}
		\begin{tikzpicture}[>=Stealth]
            \fill[blue!20, opacity=0.5] (-1,-1,0) -- (-1,1,0) -- (1,1,0) -- (1,-1,0) -- cycle;
            \fill[green!20, opacity=0.5] (0,-1,-1) -- (0,-1,1) -- (0,1,1) -- (0,1,-1) -- cycle;
            \fill[red!20, opacity=0.5] (-1,0,-1) -- (-1,0,1) -- (1,0,1) -- (1,0,-1) -- cycle;
            \draw (-1,-1,-1) -- (-1,-1,1) -- (-1,1,1) -- (-1,1,-1) -- cycle;
            \draw (1,-1,-1) -- (1,-1,1) -- (1,1,1) -- (1,1,-1) -- cycle;
            \draw (-1,-1,-1) -- (1,-1,-1);
            \draw (-1,-1,1) -- (1,-1,1);
            \draw (-1,1,-1) -- (1,1,-1);
            \draw (-1,1,1) -- (1,1,1);
            \foreach \x/\y/\z in {1/0/0, 0/1/0, 0/0/1, -1/0/0, 0/-1/0, 0/0/-1, 1/1/0, 1/-1/0, -1/1/0, -1/-1/0, 1/0/1, 1/0/-1, -1/0/1, -1/0/-1, 0/1/1, 0/1/-1, 0/-1/1, 0/-1/-1} {
              \draw[->, thick] (0,0,0) -- (\x,\y,\z);
            }
        \end{tikzpicture}
    \end{subfigure}
    \begin{subfigure}{0.3\textwidth}
		\begin{tikzpicture}[>=Stealth]
            \fill[blue!20, opacity=0.5] (-1,-1,0) -- (-1,1,0) -- (1,1,0) -- (1,-1,0) -- cycle;
            \fill[green!20, opacity=0.5] (0,-1,-1) -- (0,-1,1) -- (0,1,1) -- (0,1,-1) -- cycle;
            \fill[red!20, opacity=0.5] (-1,0,-1) -- (-1,0,1) -- (1,0,1) -- (1,0,-1) -- cycle;
            \draw (-1,-1,-1) -- (-1,-1,1) -- (-1,1,1) -- (-1,1,-1) -- cycle;
            \draw (1,-1,-1) -- (1,-1,1) -- (1,1,1) -- (1,1,-1) -- cycle;
            \draw (-1,-1,-1) -- (1,-1,-1);
            \draw (-1,-1,1) -- (1,-1,1);
            \draw (-1,1,-1) -- (1,1,-1);
            \draw (-1,1,1) -- (1,1,1);
            \foreach \x/\y/\z in {1/0/0, 0/1/0, 0/0/1, -1/0/0, 0/-1/0, 0/0/-1, 1/1/0, 1/-1/0, -1/1/0, -1/-1/0, 1/0/1, 1/0/-1, -1/0/1, -1/0/-1, 0/1/1, 0/1/-1, 0/-1/1, 0/-1/-1, 1/1/1, 1/1/-1, 1/-1/1, 1/-1/-1, -1/1/1, -1/1/-1, -1/-1/1, -1/-1/-1} {
              \draw[->, thick] (0,0,0) -- (\x,\y,\z);
            }
        \end{tikzpicture}
    \end{subfigure}
    \caption{D2Q9, D3Q19, and D3Q27 lattices containing $q=9, 19, 27$ vectors respectively.}
	\label{fig:lattice}
\end{figure}
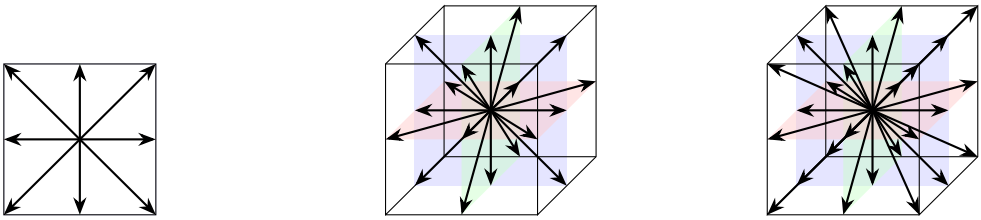

The dynamics of an evolving flow field may be represented at mesoscopic scales using the LBM equations. These inherently time-dependent and discrete equations, govern the spatial and temporal behaviour of a set of \emph{velocity distribution functions} or $f_i(\boldsymbol{x},t)$, through a \emph{collide-and-stream} algorithm. In its most general form, the LBM equations may be formulated based on a general collision operator $\mathscr{C}$ as:
\begin{alignat}{2}
	&\mbox{Collision:} \qquad && f_i^\ast(\boldsymbol{x}, t)  = \mathscr{C}(f_i(\boldsymbol{x}, t)) \label{eq:collision} \\
	&\mbox{Streaming:} \qquad && f_i(\boldsymbol{x} + \Delta\boldsymbol{x} , t + \Delta t) = f_i^\ast(\boldsymbol{x}, t) \label{eq:streaming}
\end{alignat}
Notice that the subscript `$i$' indexes the discrete lattice directions along which the above collide-and-stream operator applies. Each discrete cell situated at $ \boldsymbol{x}$, with spatial spacing $\Delta\boldsymbol{x} = \boldsymbol{c}_i\Delta t$, has a `lattice' structure, formally denoted by a set of $q$ vectors $\boldsymbol{c}_i = \{\boldsymbol{c}_1, \ldots, \boldsymbol{c}_q\}$ that are visually illustrated in Figure~\ref{fig:lattice} for 2D and 3D settings.

The collision operator in XLB may be of any form including (but not limited to) the single-relaxation model due to Bhatnagar-Gross-Krook (BGK), the multi-relaxation time (MRT) method (see \cite{Kruger:2016:book}) based on any form of the moment space, or the more advanced collision operators such as the cumulant collision \cite{Geier:2015:cumulant}, recursive regularized \cite{coreixas2017recursive} or the multi-relaxation entropic model (also known as the `KBC' model) \cite{Karlin:2014:gibbs}. 
Without loss of generality, we only present the classic BGK model here,
\begin{equation}
    \mathscr{C}_{\text{BGK}} := f^\ast_i(\boldsymbol{x}, t) = f_i(\boldsymbol{x}, t)  - \frac{\Delta t}{\tau} \left[ f_i(\boldsymbol{x}, t) - f_i^{eq}(\boldsymbol{x},t) \right]
      \label{eq:LBM-BGK}
\end{equation}
which indicates the relaxation of $f_i$ with respect to an \emph{equilibrium} state, $f_i^{\mathit{eq}}$, with a timescale $\Delta t/ \tau$ where $\Delta t$ is the discrete time step and $\tau$ is related to the total kinematic viscosity $\nu$ of the fluid as,
\begin{equation}
\tau = \frac{\nu}{c_s^2} + \frac{\Delta t}{2}.
\label{eq:tau}
\end{equation}
in which $c_s$ is the constant speed of sound given by $c^2_s = (1/3) (\Delta\boldsymbol{x} / \Delta t)^2$.

Independent of the collision model, the equilibrium distribution function $f^{eq}_i$ (in its quadratic form) is defined as,
\begin{equation}
        f_i^{\mathit{eq}}(\boldsymbol{x}, t) = 
        w_i \rho(\boldsymbol{x}, t) \left\lbrace  1 + \frac{\boldsymbol{c}_i \cdot \boldsymbol{u}(\boldsymbol{x}, t)}{c_s^2} + \frac{(\boldsymbol{c}_i \cdot \boldsymbol{u}(\boldsymbol{x}, t))^2}{2c_s^4} - \frac{\| \boldsymbol{u}(\boldsymbol{x}, t) \|^2}{2c_s^2} \right\rbrace 
        \label{eq:feq}
\end{equation}
where $w_i$ are weights associated with each lattice direction $\boldsymbol{c}_i$. In addition, the macroscopic state variables namely the fluid density $\rho(\bm{x},t)$, velocity $\bm{u}(\bm{x},t)$ and pressure $p(\bm{x},t)$ are formally derived from $f_i$ as,
\begin{alignat}{2}
	\mbox{density:}\quad &&  \rho(\boldsymbol{x}, t) &= \sum_i f_i(\boldsymbol{x}, t) \label{eq:density} \\
	\mbox{velocity:}\quad && \bm{u}(\boldsymbol{x}, t)&= \frac{1}{\rho(\boldsymbol{x}, t)} \sum_i \boldsymbol{c}_i f_i(\boldsymbol{x}, t) \label{eq:velocity}\\
    \mbox{pressure:}\quad && p(\boldsymbol{x}, t)&= c_s^2 \rho(\boldsymbol{x}, t) \label{eq:pressure}
\end{alignat}

At the boundaries of the computational domain, where $\boldsymbol{x}+\boldsymbol{c}_i\Delta t$ would leave the assigned computational domain, it is necessary to impose boundary conditions such that the desired physical system can be simulated reasonably (see Section~\ref{sec:boundary_conditions} for further details). Again there exists a multitude of options in the literature and all can be efficiently and easily implemented in XLB. For an updated list of available schemes in XLB the reader is referred to the online repository.

%% file: tex/impl.tex
\section{Implementation details}
\label{sec:impl}
The following sections will briefly explain the integration of JAX within XLB, highlighting how its features are leveraged to achieve high-performance computing in LBM. It is important to note that the XLB library is continually evolving, with frequent updates and enhancements. For the most up-to-date information, readers are advised to consult the XLB repository directly. As such, this section aims to provide a general overview of its structure and functionalities, considering that the specifics may evolve beyond the scope of this document at the time of the reader's perusal.

\subsection{A brief introduction to JAX}
\label{sec:jax}
JAX \cite{jax2018github}, originally developed by Google, is a high-performance numerical computing library that offers significant advantages for scientific and engineering computations, particularly in the field of machine learning. It forms part of an extensive ecosystem of machine learning libraries such as Flax~\cite{flax2020github}, Equinox~\cite{kidger2021equinox}, and Optax~\cite{deepmind2020jax}, making it adaptable for various use cases. 

At the core of JAX's functionality is its ability to transform Python functions. As a result, it can automatically differentiate, compile to GPU/TPU, and vectorize code, making it highly versatile and efficient for large-scale computations. The \texttt{jax.numpy} module provides standard NumPy functionality but with the added benefits of JAX's transformations, offering a similar interface while enabling acceleration on GPUs and TPUs. This makes it familiar and accessible to users with a background in NumPy, while offering considerable performance gains.

One of the key features of JAX is its JIT compilation, which compiles Python functions into highly optimized machine code. This feature significantly boosts performance, especially in computational loops or repeated operations. JAX also supports automatic vectorization with \texttt{vmap}, allowing for batch processing of functions for efficiency. The convenience of JAX's NumPy interface, coupled with the efficiency of JIT compilation, positions XLB as a suitable choice for a wide range of computational tasks, especially for physics-based machine learning.

XLB incorporates object-oriented design for certain functionalities. To ensure compatibility with JAX, a functional programming library, XLB leverages the \texttt{partial} decorator. This decorator freezes specific parameters (like the \texttt{self} argument in classes) as static during JIT compilation process. This allows JIT-compiled functions to work within XLB's object-oriented structure while avoiding the side effects that arise from JAX's preference for stateless functions~\cite{jax2018github}.

\subsection{Lattice definitions}
\label{sec:lattice_definition}
The library defines a hierarchy of classes to encapsulate various lattice configurations. The architecture consists of a parent class, denoted as \texttt{Lattice}, and specific subclasses for common lattice configurations such as \texttt{LatticeD2Q9}, \texttt{LatticeD3Q19}, and \texttt{LatticeD3Q27}. The parent class defines the core attributes and methods that are common to all lattice configurations, such as moments, weights, velocity vectors, and more. The subclasses inherit these attributes and methods, further specialize them by introducing modifications tailored to their specific lattice configurations. See Figure~\ref{fig:lattice} for a visual illustration of these lattice structures.

\subsection{Simulation domain and array representation}
\label{sec:simulation_domain}
The simulation domain is represented as a grid defined in Cartesian coordinates, characterized by the number of grid points along each axis: \texttt{nx}, \texttt{ny}, and \texttt{nz}. These parameters must be provided during the initialization process. For 2D simulations, \texttt{nz} should be set to zero. The computational domain is discretized using uniform equidistant tiles/cells in 2D and 3D.

XLB utilizes \texttt{jax.Array} and a sharding strategy to distribute the computational load across multiple devices. The dimensions of each distributed array in XLB are ordered in an array-of-structure format as (\texttt{x}, \texttt{y}, \texttt{z}, \texttt{cardinality}) for 3D simulations, and (\texttt{x}, \texttt{y}, \texttt{cardinality}) for 2D simulations. For example, the array associated with the distribution functions (or `populations') $f_i$ for a D3Q19 lattice are defined as an array with dimensions (\texttt{nx}, \texttt{ny}, \texttt{nz}, \texttt{19}), while the density array has dimensions of (\texttt{nx}, \texttt{ny}, \texttt{nz}, \texttt{1}). 

\begin{figure}[H]
  \centering
  \includegraphics[width=0.8\textwidth]{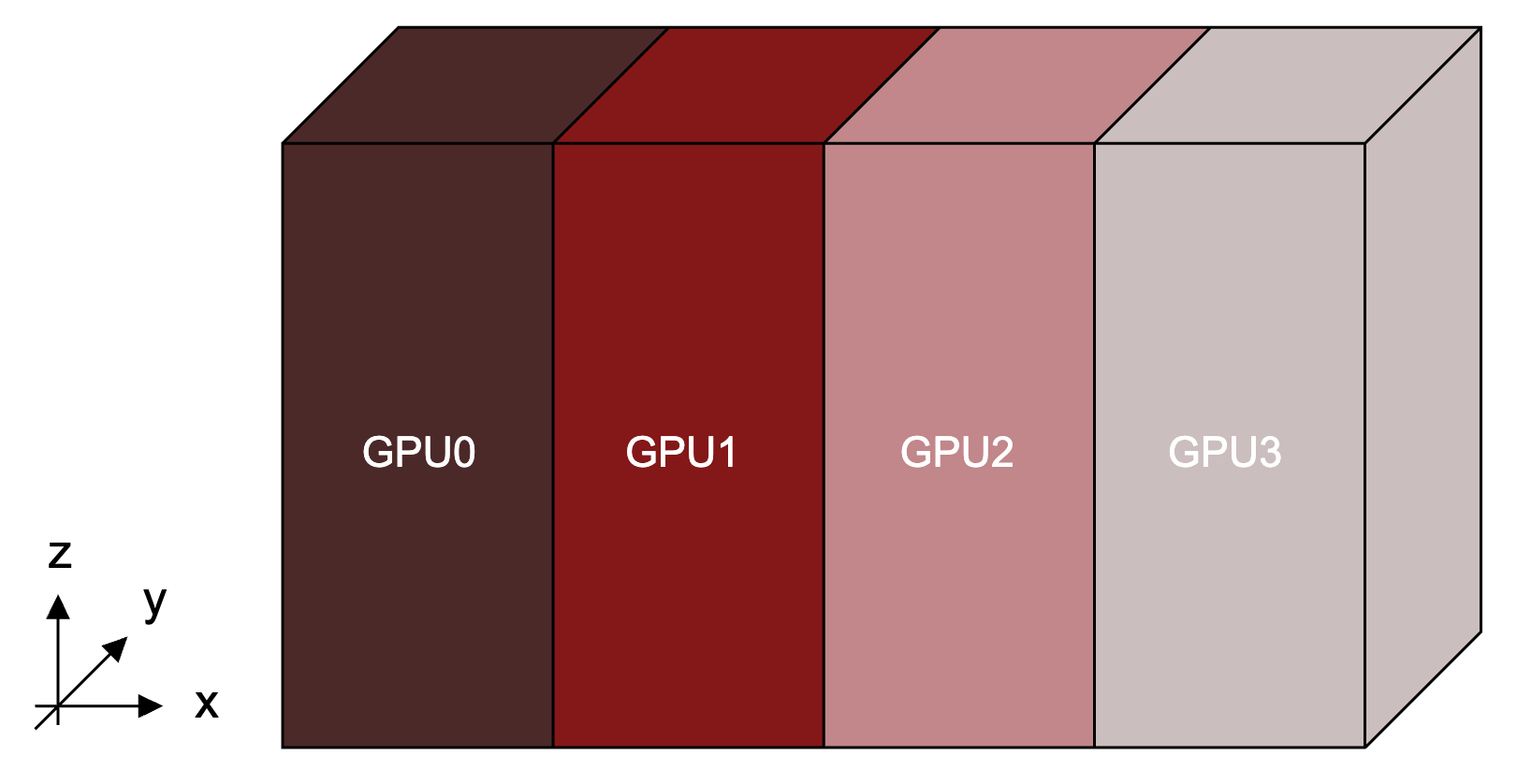}
  \caption{Illustration of the array sharding strategy employed by XLB for efficient computational load distribution across multiple devices. The strategy requires that the number of grid points along the \texttt{x}-axis to be divisible by the number of available devices. This requirement is automatically handled by the library based on the user's choice for \texttt{nx}.}\label{fig:sharding}
\end{figure}

Utilizing uniform axis conventions for all fields allows the library to execute automated sharding along the \texttt{x}-axis, a process visualized in Figure~\ref{fig:sharding}. For this purpose, JAX requires the number of grid points along the \texttt{x}-axis, \texttt{nx}, to be divisible by the number of available devices \texttt{nDevices} (e.g., number of GPUs). If \texttt{nx} is not divisible by \texttt{nDevices}, the library automatically increases \texttt{nx} to the nearest higher multiple of \texttt{nDevices}. While the current implementation focuses on \texttt{x}-axis sharding, expanding to multi-dimensional sharding could further optimize performance, especially when scaling to larger clusters, addressing the increased communication overhead that will be noted in Section~\ref{sec:multi_node_scaling}.

Our approach shards XLB arrays across multiple GPUs, enabling global operations on the entire dataset without requiring separate computations. While efficient and easy to use, some operations may still benefit from explicit communication between GPUs. JAX provides protocols for this targeted communications, allowing us to fine-tune data distribution and processing during complex distributed operations. Section~\ref{sec:streaming} of the manuscript details their use for the streaming operation.

Distributed arrays are initialized across multiple devices using a specialized method within the XLB library. This approach is essential for large-scale simulations, especially when the array size surpasses the memory capabilities of a single GPU. In such cases, initializing the entire array on one device and then distributing it is not practical. The library addresses this challenge by incorporating a method that employs sharding constraints. These constraints, guided by JAX's sharding functionality, ensure that the JIT compiler allocates only the necessary portion of an array that each device is responsible for. This method effectively manages memory across multiple devices, optimizing the allocation and utilization of resources in distributed computing environments.

\subsection{Streaming operation and distributed computing}
\label{sec:streaming}
The streaming operation (see equation~\ref{eq:streaming}) is typically the only non-local operation in LBM that demands communication across devices, and therefore it warrants special consideration in a multi-device setup. To enable this, the XLB library employs a suite of functions: \texttt{streaming\_m}, \texttt{streaming\_p}, and \texttt{streaming\_i}.

The primary function, \texttt{streaming\_m}, orchestrates the streaming processes and enables the inter-process communication in a multi-device context. This function isolates specific portions of populations from the left and right boundary slices of distribution arrays on each device. These isolated portions are then communicated to adjacent processes.\ \texttt{streaming\_p} function handles streaming operations on a partitioned sub-domain within a single device. It leverages the \texttt{vmap} operation from the JAX library to carry out vectorized computations across all lattice directions. The \texttt{streaming\_i} function executes individual streaming operations for each directional index $i$. This is achieved using the \texttt{jnp.roll} function from the JAX library, which shifts distribution function values in the direction specified by each discrete velocity vector $\boldsymbol{c}_i$ (see Figure~\ref{fig:lattice} for an illustration of these vectors for various lattice structures).

The \texttt{jnp.roll} function effectively simulates the streaming step by moving the distribution function values from one lattice cell to the next. The shift behavior of \texttt{jnp.roll} is toroidal, allowing populations $f_i$ at the domain boundaries to traverse to the opposite end, mirroring the dynamics in toroidal or periodic manner, effectively imposing periodic boundary conditions in all directions. When a population at the edge of one sub-domain (on one device/GPU) is shifted, it must appear at the converse boundary of the adjacent sub-domain to maintain a consistent toroidal shift across the entire domain (on all devices). However, this ``cross-boundary'' shift is not automatically handled by \texttt{jnp.roll}; in its standard operation, \texttt{jnp.roll} shifts elements along a specified axis only within a single sub-domain. To circumvent this limitation, the \texttt{streaming\_m} function employs explicit permute communication among the computing devices. A pre-defined set of permutations establishes which devices are adjacent, enabling the \texttt{lax.ppermute} function to transfer slices of populations to adjacent sub-domains. These slices are communicated to the right and left sub-domain neighbors using the \texttt{send\_right} and \texttt{send\_left} functions, respectively. The received data is then placed at the appropriate indices within the receiving sub-domain. This mechanism ensures the continuity of the toroidal shift pattern across multiple devices.

\begin{figure}[H]
    \centering
    \includegraphics[width=0.8\textwidth]{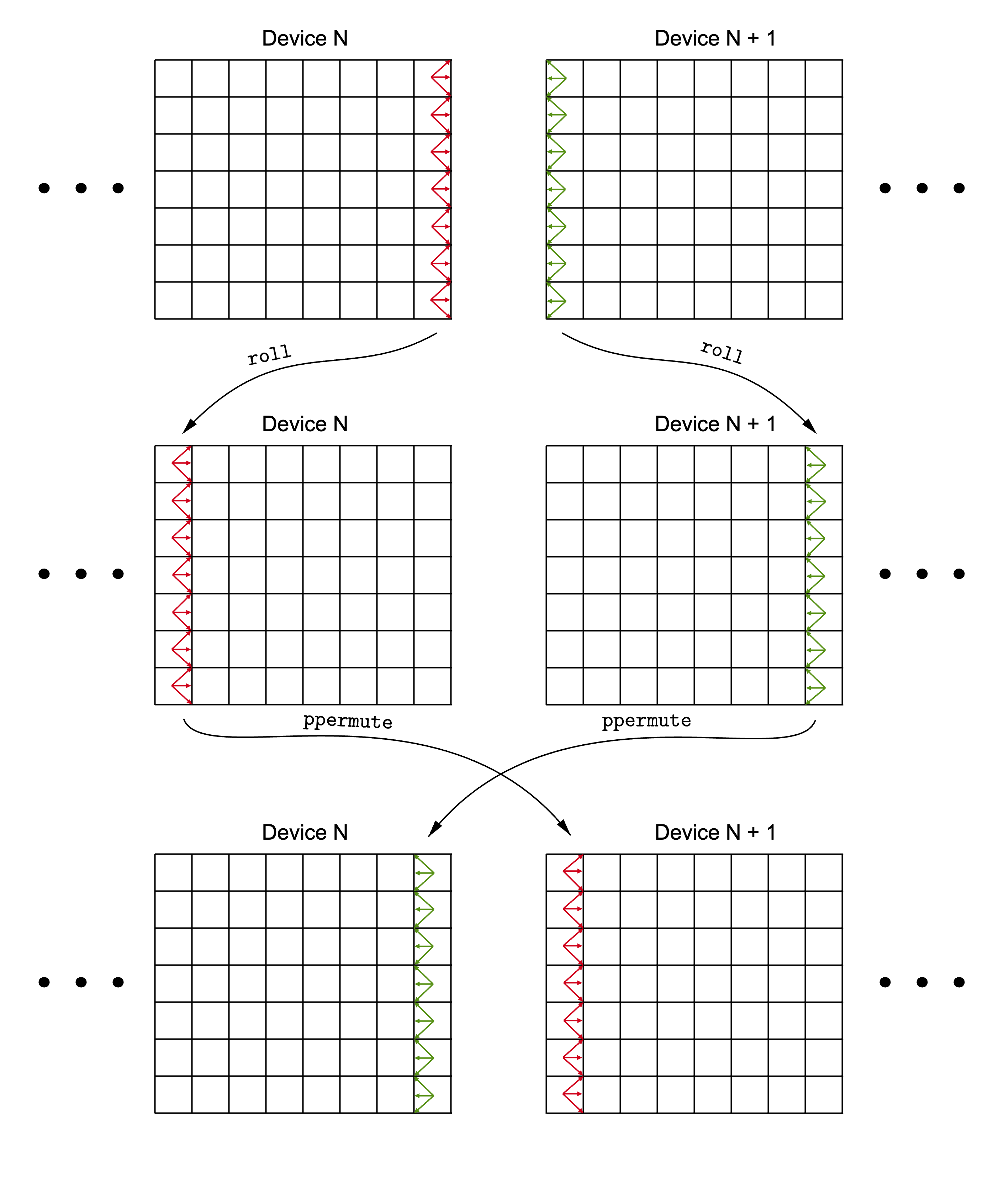}
    \caption{The streaming operation in distributed computing, showcasing how population arrays are communicated between devices in XLB library. }\label{fig:streaming}
\end{figure}

To illustrate these streaming communications, Figure~\ref{fig:streaming} provides a visual guide. First, the \texttt{jnp.roll} function shifts populations in the direction of discrete velocities $\boldsymbol{c}_i$ (shown schematically for the \texttt{D2Q9} lattice), handling the populations at the boundaries of each sub-domain in a toroidal manner. Subsequently, slices requiring communication to adjacent sub-domains are identified and permuted using \texttt{send\_right} and \texttt{send\_left} functions. The incoming data are assigned to the designated indices within the recipient sub-domain. This communication strategy effectively extends the toroidal shift behavior of \texttt{jnp.roll} across multiple devices, without requiring halo cells to be allocated for the multi-device communication.

JAX's \texttt{shard\_map} is used to specify manually how parallel communications are orchestrated during streaming. This design choice enhances code readability; it allows for a more straightforward representation of arrays in their original forms, where all arrays retain their original, physically intuitive dimensions outside the \texttt{shard\_map} function, simplifying code structure and reducing the need for complex array manipulations. This is in contrast to using JAX's \texttt{pmap} (parallel map for collective operations), which necessitates adding an extra array axis to represent device numbers, potentially complicating the codebase.

\subsection{Boundary conditions}
\label{sec:boundary_conditions}
XLB includes methods for both setting up boundary attributes and applying boundary conditions (BCs). Specialized BCs inherit from a base class and override methods and variables as needed for their specific implementation.

Certain BCs, like equilibrium and full-way bounce-back, merely require identification of the boundary cells for their application. In contrast, other BCs, such as Zou-He~\cite{zou1997pressure}, regularized \cite{Latt:2008:regularized} or half-way bounce-back, demand additional supplementary data. This may entail knowing the types of the surrounding cells, the distance to the solid mesh, the direction of unknown populations requiring reconstruction by the BC scheme, the boundary's normal vector, and conditions specifying whether the BC must be applied to the adjacent fluid cell. To facilitate this, an initialization step generates a boolean mask array similar to the size and cardinality of the population array $f_i$. In this array, for each direction $\boldsymbol{c}_i$, the value is set to \texttt{True} if $-\boldsymbol{c}_i$ points to solid neighbors (conceptually equivalent to directions where $f_i$ is streamed from solid neighbors) and conversely to \texttt{False} if $-\boldsymbol{c}_i$ points to a fluid neighbor (conceptually equivalent to directions where $f_i$ is streamed from fluid neighbors). 

\begin{figure}[H]
    \centering
    \includegraphics[width=\textwidth]{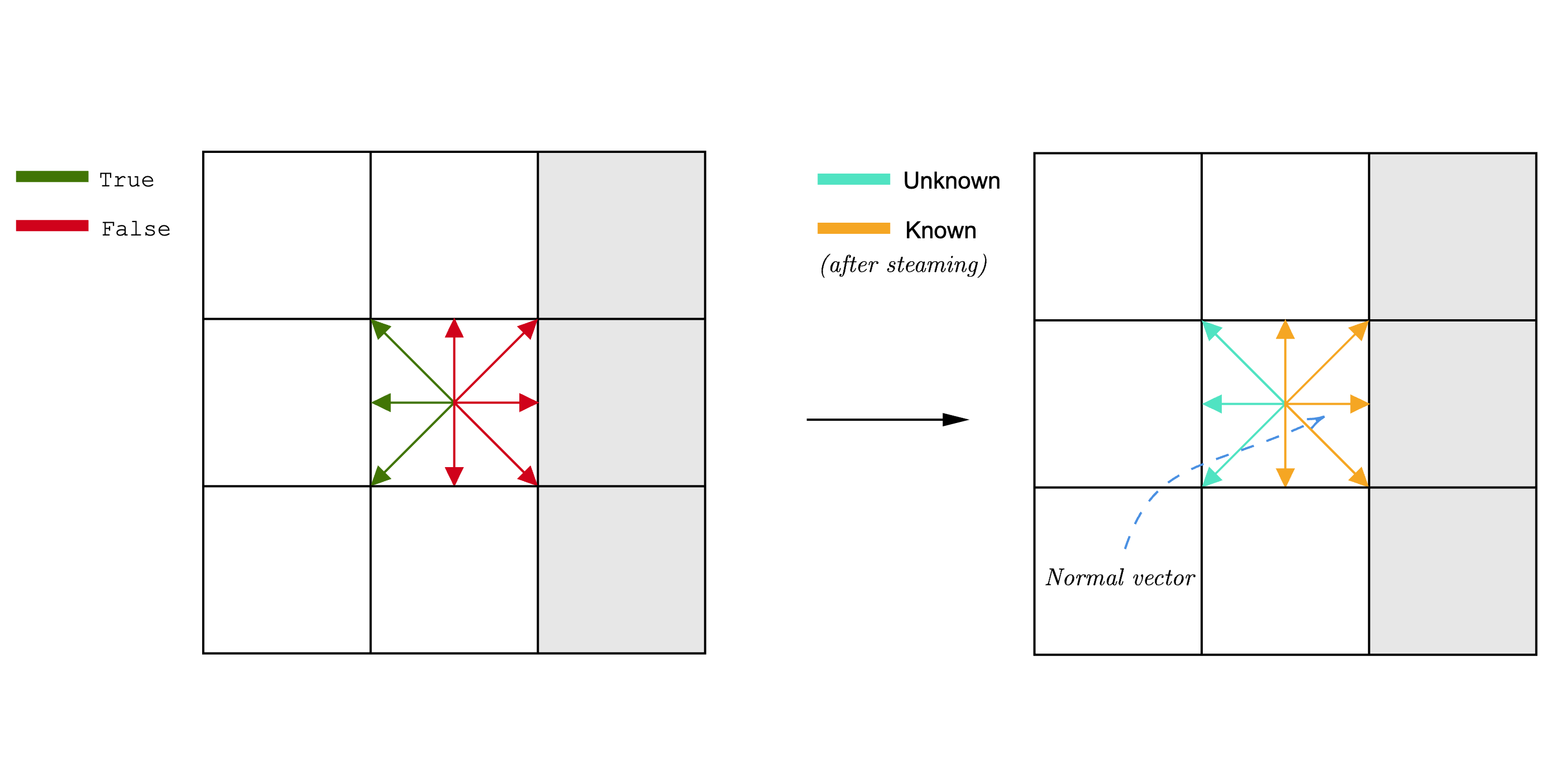}
    \caption{Visual representation of utilizing a mask array for identifying normal vector and distinguishing known and unknown boundary populations post-streaming.}\label{fig:mask}
\end{figure}

The process of constructing the boolean mask array involves a few steps that are described here. First, an array of \texttt{False} with dimensions of (\texttt{nx}, \texttt{ny}, \texttt{nz}, \texttt{q}) is extended along $x, y$ and $z$ directions to accommodate halo layers on the periphery of the computational domain. The number of halo layers in $x$ is chosen as a multiple of \texttt{nDevices} to enable array sharding. Then, in order to delineate between `exterior' and `interior' of the domain for the purpose of imposing the boundary conditions, values of \texttt{True} are assigned to these peripheral halo cells as well as the internal solid cells defined through the BC constructs. A key step in constructing the above mask array is to perform a single streaming on the extended array to shift the boolean content in such a way that leads to an easy realization of \emph{missing} (associated with \texttt{True}) and \emph{known} (associated with \texttt{False}) lattice directions for the boundary cells. A visual representation that shows the role of the mask array in identifying the normal vector as well as distinguishing between known and unknown populations at the boundary after the streaming step is depicted in Figure~\ref{fig:mask}.

The boundary conditions rely on the mask array to modify the allocated boundary cells appropriately (if required by the boundary type) and to generate any supplementary information required for implementing the BC scheme such as the normal direction or the momentum flux.

\subsection{LBM step}
\label{sec:lbm_step}
The \texttt{step} method orchestrates the key stages of the LBM algorithm, namely the collision, boundary condition application, and streaming operations. One LBM step includes both collision and streaming operations, each followed by the application of boundary conditions. Depending on the boundary type, some schemes are imposed after collision, while others after streaming.

The \texttt{step} method effectively simulates one time step in the fluid flow. Optionally, it can return the post-streaming and post-collision states for additional analysis, such as that required for calculating the lift and drag forces.

Leveraging JAX's automatic differentiation capabilities, users can conveniently compute the gradient of a single LBM step with respect to its inputs. By applying the \texttt{jax.grad} function to the \texttt{step} method, one can directly obtain the gradients needed for sensitivity analysis, optimization, and machine learning -- all without manual differentiation.

\subsection{Mixed precision, distributed checkpointing, and I/O capabilities}
\label{sec:distributed_checkpointing_mixed_precision}
The XLB library stands out with its support for mixed-precision computation. This allows users to tailor precision levels for computation and storage independently. For instance, a setting like \texttt{f64/f32} indicates that the computation is performed in double precision (f64) while storage is handled in single precision (f32). Mixed-precision offers a balance of accuracy and efficiency: reduced memory usage enables larger simulations and potential performance gains, especially on modern GPUs due to their superior performance in lower precision arithmetic.

XLB also features robust checkpointing for long-running simulations. Users can save simulation states at intervals, facilitating recovery from system failures or allowing simulations to be paused and resumed at will.  XLB's distributed checkpointing system is optimized for simulations running across multiple devices or nodes.

The XLB library supports both Binary and ASCII VTK output formats through the PyVista library~\cite{sullivan2019pyvista}. Integration with PhantomGaze~\cite{PhantomGaze} enables advanced, GPU-accelerated in-situ visualizations. This integration is streamlined by DLPack, which allows tensor data exchange between XLB and PhantomGaze directly on the GPU, eliminating traditional I/O bottlenecks. For a given simulation, users can customize the frequency of I/O and extract specific quantities of interest, such as velocity, pressure, lift, drag, energy spectrum, strain rate tensor, viscous dissipation and more.

We are actively expanding the capabilities of the XLB library. Please check out our repository for the latest updates and features.

%% file: tex/benchmarks.tex
\section{Benchmarks}
\label{sec:benchmarks}
In this section, we aim to demonstrate the robustness and accuracy of XLB for simulating laminar and turbulent flows under different set of initial and boundary conditions. For this purpose we will discuss four standard benchmark problems, namely (i) the Taylor-Green Vortex in 2D, (ii), the lid-driven cavity in 3D (iii) open flow over a 2D cylinder and (iv) turbulent flow inside a 3D channel. As will be showcased for each example, a distinct set of XLB capabilities are targeted and verified in each case.

\subsection{Taylor-Green vortex in 2D}
\label{sec:tg}
%
The Taylor-Green Vortex problem consists of a freely decaying flow in a periodic setting that is initialized by a particular distribution of velocity and pressure fields. This example is an interesting benchmark problem in 2D as the analytical solution to the governing conservation laws may be derived to take the following form,

\begin{eqnarray}
    u_{th}(\boldsymbol{x}, t) &=& U_0 \sin(k_x x) \cos(k_y y) \exp\left\lbrace -\frac{t}{t_d} \right\rbrace \nonumber \\
    v_{th}(\boldsymbol{x}, t) &=& -U_0 \cos(k_x x) \sin(k_y y) \exp\left\lbrace -\frac{t}{t_d} \right\rbrace \nonumber \\
    p_{th}(\boldsymbol{x}, t) &=& 1 -\frac{\rho_0 U_0^2}{4} \left[ \cos(2 k_x x) + \cos(2 k_y y) \right] \exp\left\lbrace -2 \frac{t}{t_d}\right\rbrace
    \label{eqn:tg:analytical}
\end{eqnarray}
in which $k_x = 2\pi / n_x$ and $k_y = 2\pi / n_y$ with $n_x$ and $n_y$ indicating the number of cells in $x$ and $y$ directions, respectively. Also, $t_d$ denotes a diffusive timescale defined as $t_d = \left[2 \nu ( k_x^2 + k_y^2) \right]^{-1}$.

To compare simulation results with the above analytical formulations, we define the relative $L_2$-errors for velocity and density fields as,
\begin{equation}
    \epsilon_u = \frac{ \displaystyle \iint\left[ (\bm{u}(\boldsymbol{x}, t_{f}) - \bm{u}_{th}(\boldsymbol{x}, t_f) \right]^2 \, d\bm{x} }{\displaystyle \iint ||\bm{u}_{th}(\boldsymbol{x}, t_{f})||^2 \, d\bm{x}}, \qquad 
    \epsilon_\rho = \frac{ \displaystyle \iint \left[\rho(\boldsymbol{x}, t_{f}) - \rho_{th}(\boldsymbol{x}, t_{f}) \right]^2 \, d\bm{x}}{\displaystyle \iint \rho_{th}^2(\boldsymbol{x}, t_{f}) \, d\bm{x}}
    \label{eqn:error_tg}
\end{equation} 
where $t_f$ indicates the final time of simulation and $\rho_{th} = p_{th}/3$. 

\begin{figure}[H]
	\centering
	\includegraphics[width=0.8\textwidth]{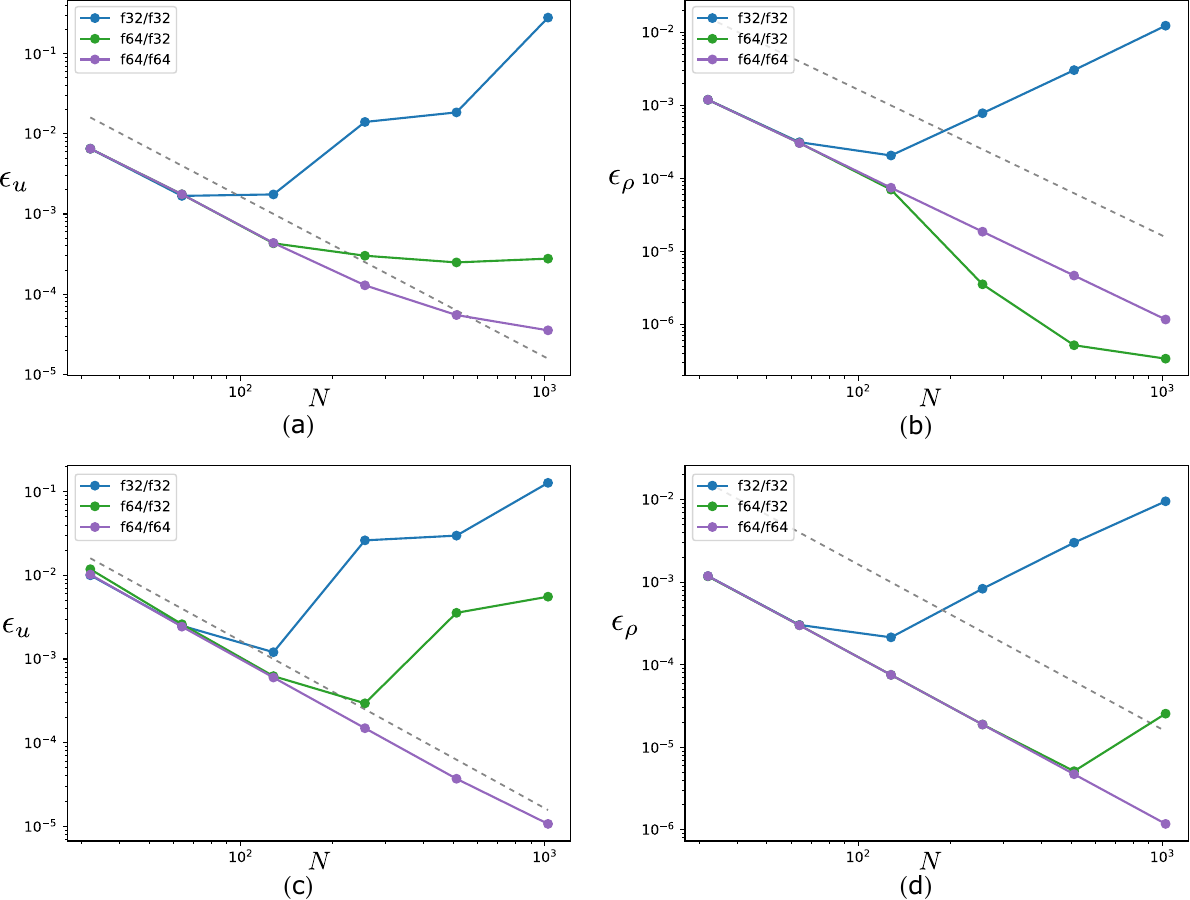}
	\caption{\label{fig:tg} Illustrating the variation of $\epsilon_{u}$ and $\epsilon_{\rho}$~\eqref{eqn:error_tg} associated with the 2D Taylor-Green Vortex problem with increasing resolution for both BGK (a, b) and KBC (c, d) collision models. The results are obtained with various `computation/storage' pairs where for instance \texttt{f64/f32} represents double precision accuracy for computations with single precision accuracy for storage.}
\end{figure}

Figure~\ref{fig:tg} illustrates $\epsilon_u$ (in panels a, c) and $\epsilon_\rho$ (in panels b, d) for a range of resolutions $N= \lbrace 32, \ldots, 1024 \rbrace$ using either the BGK (in panels a, b) or the KBC (in panels c, d) models as the collision operator. The reported simulation results were conducted using various mixed precision `computation/storage' pairs where \texttt{f64} and \texttt{f32} indicate double and single precision, respectively. For all cases, we used the D2Q9 lattice and periodic boundary conditions (recall that by default if no boundary condition is imposed, the streaming operation in XLB imposes periodic boundary conditions by construction). All the numerical results were obtained at $t_f = 0.05 t_d$ using a flow with $Re= 1/\nu = 1600$. We set $\Delta x = 1/N$ and $\Delta t/ \Delta x = 0.04$ at $N=32$ where $\Delta t$ was decreased proportionally as $N$ was increased. The distribution functions $f_i$ were initialized based on the method of~\cite{Mei:2006:ic} using $\bm{u}_{th}(\bm{x},t=0)$ as per~\eqref{eqn:tg:analytical}. 

The expected second order of accuracy (i.e. $\epsilon_u \propto N^{-2}$ and $\epsilon_\rho \propto N^{-2}$) is obtained for both collision models using \texttt{f64/f64} precision pair. It is worth mentioning that $\epsilon_u$ shows a more robust convergence based on the KBC model up to $N=1024$ while the convergence of $\epsilon_u$ degrades for the BGK model beyond $N=256$ for \texttt{f64/f64}. We may attribute this to the chosen time step size or the number of iterations used for the initialization process. If the storage precision is reduced to `f32' (i.e. \texttt{f64/f32} cases), the second order accuracy is preserved up to a lower resolution beyond which both $\epsilon_u$ and $\epsilon_\rho$ deteriorate as $N$ increases. This is due to the fact that the results become cluttered by numerical rounding errors and hence are not improved with increasing the resolution. This exact deterioration in accuracy occurs more vividly for even lower resolutions as the computational precision is also reduced to \texttt{f32} (i.e. \texttt{f32/f32} cases) leading to erroneous results for higher resolutions. Consequently, LBM results with \texttt{f32/f32} are not reliable unless the distribution functions $f_i$ values are renormalized carefully to be centered around zero~\cite{lehmann2022accuracy} (this feature is not currently available in XLB).

We wish to highlight that for new GPU architectures like Amper and Hopper, dot product computations (e.g., in the equilibrium function) utilize TensorFloat32 (\texttt{TF32}) rather than float by default. This behavior can be altered by \texttt{jax.default\_matmul\_precision}. Nonetheless, in the scenarios depicted in Figure~\ref{fig:tg}, we did not detect any notable differences in the outcomes when employing \texttt{TF32} as opposed to \texttt{f32}.

\subsection{Lid-driven cavity in 3D}
Here we investigate the 3D flow inside a lid-driven cavity to further verify XLB especially under steady-state, unsteady and turbulent scenarios with non-periodic boundaries. Namely, Dirichlet boundary conditions are employed to assign no-slip condition on all side walls except for the top lid which moves horizontally with a prescribed velocity of $\bm{u}(x,y,z=L/2) = (U, 0, 0)$. We consider a box of size $L^3$ containing an incompressible Newtonian fluid with viscosity $\nu$ that is characterized by a Reynolds number defined as $Re = U L / \nu$. Each dimension is uniformly discretized using $N$ cells leading to $\Delta x = L/(N-2)$ as two cells are reserved to represent solid boundaries on either side. As in~\cite{Latt:2021:stlbm}, we fix the time-step at $\Delta t = u_{LB}/U \Delta x$ where $u_{LB}$ is related to the Mach number ($Ma$) by the speed of sound in LBM units as $u_{LB} = Ma / c_s$ and is assumed here to be constant and fixed at $u_{LB} = 0.06$ leading to $Ma \approx 0.1 \ll 1$ supporting the incompressibility requirement. Simulations are run for $n$ iterations until a final dimensionless time $t = n (U/L)\Delta t$ is reached.

\begin{figure}[H]
	\centering
	\includegraphics[width=\textwidth]{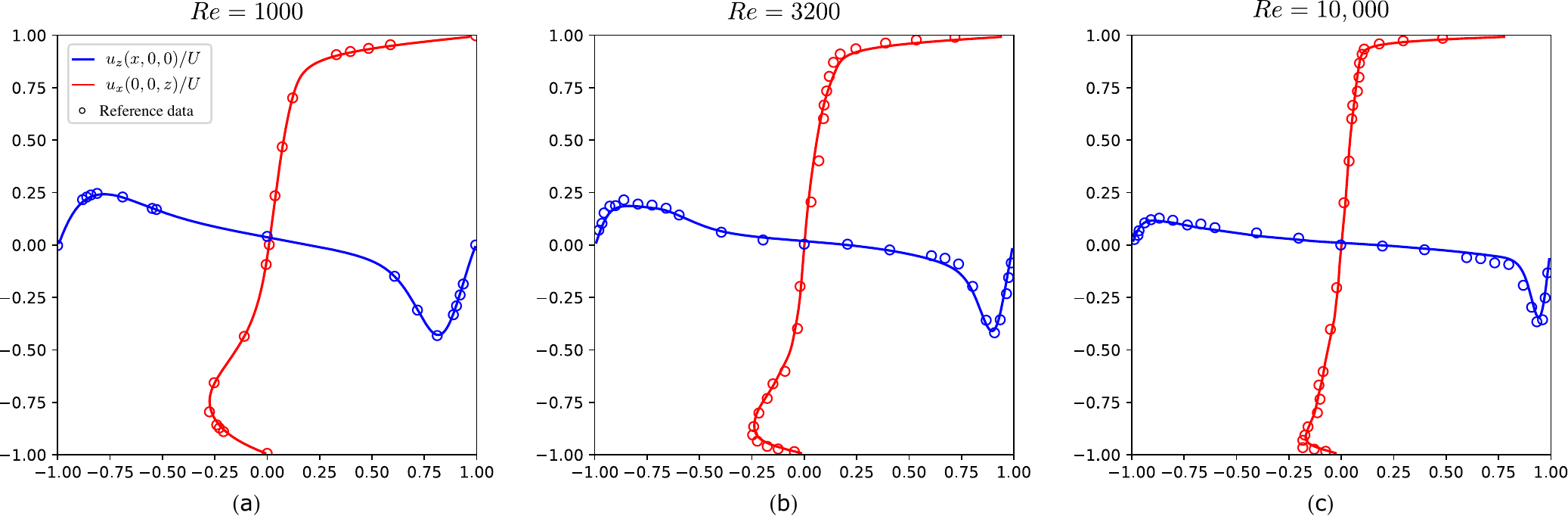}
	\caption{\label{fig:ldc} Comparing the variations of the velocity field against available reference data for 3D lid-driven cavity flow at (a) $Re=1000$, (b) $Re=3200$ and (c) $Re=10000$. In all panels the horizontal and vertical components of the velocity field ($u_x, u_z$) are probed at mid-plane of the box (i.e. at $y=0$) and are normalized w.r.t. the lid velocity (note, $u_{lid} = (U, 0, 0)$). The reference data shown by colored circles are based on (a) spectral simulation results of~\cite{Albensoeder:2005:accurate} and (b, c) experimental results of~\cite{Prasad:1989:reynolds}. The BGK collision model together with the half-way bounce-back scheme was employed in (a, b) while the KBC model together with the regularized boundary scheme was necessarily chosen for (c) at $Re=10000$ to maintain numerical stability. In all cases $N=256$, and $u_{LB} = 0.06$.}
\end{figure}

Figure~\ref{fig:ldc} demonstrates the variation of horizontal ($u_x$) and vertical $(u_z)$ components of the velocity field $\bm{u} = (u_x, u_y, u_z)$ at $Re=1000$ (steady flow), $Re=3200$ (unsteady flow) and $Re=10,000$ (turbulent flow). The velocity field is probed at the mid-plane associated with $y=0$ and normalized by the lid velocity $U$. Notice that the horizontal and vertical axes in this figure also represent the normalized (by $L/2$) distance along $x$ (for $u_z(x,0,0)/U$ on the y-axis) and along $z$ directions (for $u_x(0,0,z)/U$ on the x-axis) such that the clock-wise circulation of the flow inside the box can be visually apparent at this cross-section located at $y=0$. At $Re=3200$ and $Re=10000$, the presented results correspond with time-averaged values between $50 \leq t \leq 250$ and $150 \leq t \leq 500$ respectively similar to what has been reported in~\cite{Latt:2021:stlbm}. As shown in Figure~\ref{fig:ldc} the results produced by XLB are in excellent agreement with all the reference data obtained from~\cite{Albensoeder:2005:accurate,Prasad:1989:reynolds},

\subsection{Flow over cylinder}
\label{sec:flow_over_cylinder}
%
Another widely studied benchmark problem is that of an open flow around a circular cylinder in 2D. At sufficiently high Reynolds numbers ($Re=UD/\nu$ where $U$ is the mean speed of the incoming flow and $D$ is the diameter of the cylinder), this configuration leads to unsteady vortex shedding in the wake also known as the von K\'arm\'an vortex street. This example showcases the following additional capabilities inside XLB:
\begin{itemize}
    \item Lift and drag computations using the momentum exchange method of~\cite{Mei:2002:force} as described in~\cite{Caiazzo:2008:boundary},
    \item The no-slip boundary condition for curved geometries based on~\cite{bouzidi2001momentum},
    \item Inflow boundary condition using either the Zou-He~\cite{zou1997pressure} or the regularized approach~\cite{Latt:2008:regularized},
    \item Outflow boundary condition based on the extrapolation scheme of~\cite{Geier:2015:cumulant}.
\end{itemize}

\begin{figure}[H]
	\centering
	\includegraphics[width=\textwidth]{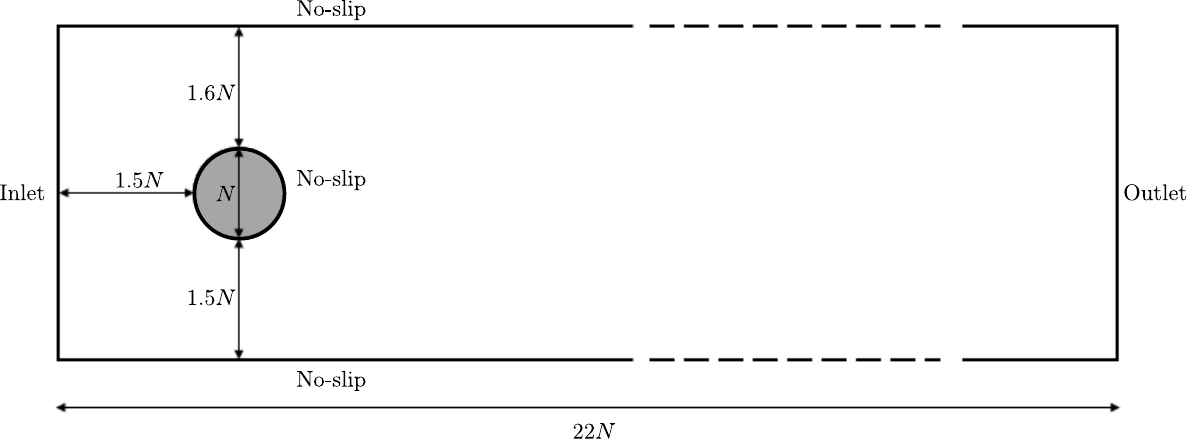}
	\caption{\label{fig:cylinder2d_setup} Illustrating the geometric setup for the flow over 2d cylinder benchmark example.}
\end{figure}

The geometric setup employed for this example is shown in Figure~\ref{fig:cylinder2d_setup} and is identical to that reported in~\cite{schafer1996benchmark} (see their figure 1). Similar to~\cite{lagrava2012advances}, a Poiseuille profile with a mean speed of $U$ is imposed at the inlet such that $Re=100$ is achieved corresponding to a laminar but unsteady condition. We discretized the computational domain using $N=80$ cells across the cylinder diameter and fixed $\Delta t = u_{LB}/U \Delta x$ assuming $u_{LB} = 0.003$ similar to~\cite{lagrava2012advances}. 

The coefficients of drag and lift are defined as,
\begin{equation}
    C_d = \frac{2F_d}{\rho U^2 D}, \qquad  C_l = \frac{2F_l}{\rho U^2 D}
\end{equation}
where $F_d$ and $F_l$ are the drag and lift forces computed as surface integrals around the cylinder. Using the BGK model and the boundary conditions outlined earlier, the following maximum coefficients were obtained after $n$ iterations corresponding with dimensionless time $t= n (U/D)\Delta t = 100$, 
\[
C^{\text{max}}_d = 3.25, \qquad C^{\text{max}}_l = 0.985.
\]
Furthermore, the vortex shedding in this example may be characterized by the Strouhal number defined as $St= f_s D / U$ in which $f_s$ is the dominant shedding frequency. Using the time series associated with $C_l(t)$ for $50 \leq t\leq 100$, we obtained $St = 0.300$. These predicted values are in good agreement with the reported range of $C_d^{\text{max}} = 3.22 - 3.24$,  $C^{\text{max}}_l = 0.99 - 1.01$ and $St = 0.295 - 0.305$ in~\cite{schafer1996benchmark} which are based on a large collection of simulations using various numerical techniques.

\subsection{Turbulent channel flow}

We now focus on a challenging and complex test associated with a continuously-forced wall-bounded turbulent flow inside a channel. This example aims to demonstrate XLB capabilities to perform accurate and efficient Large-Eddy Simulations (LES). Furthermore, this benchmark problem demonstrates how a body force can be added to the simulation using the exact-difference method of~\cite{kupershtokh2004}.

\begin{figure}[H]
	\centering
	\includegraphics[width=\textwidth]{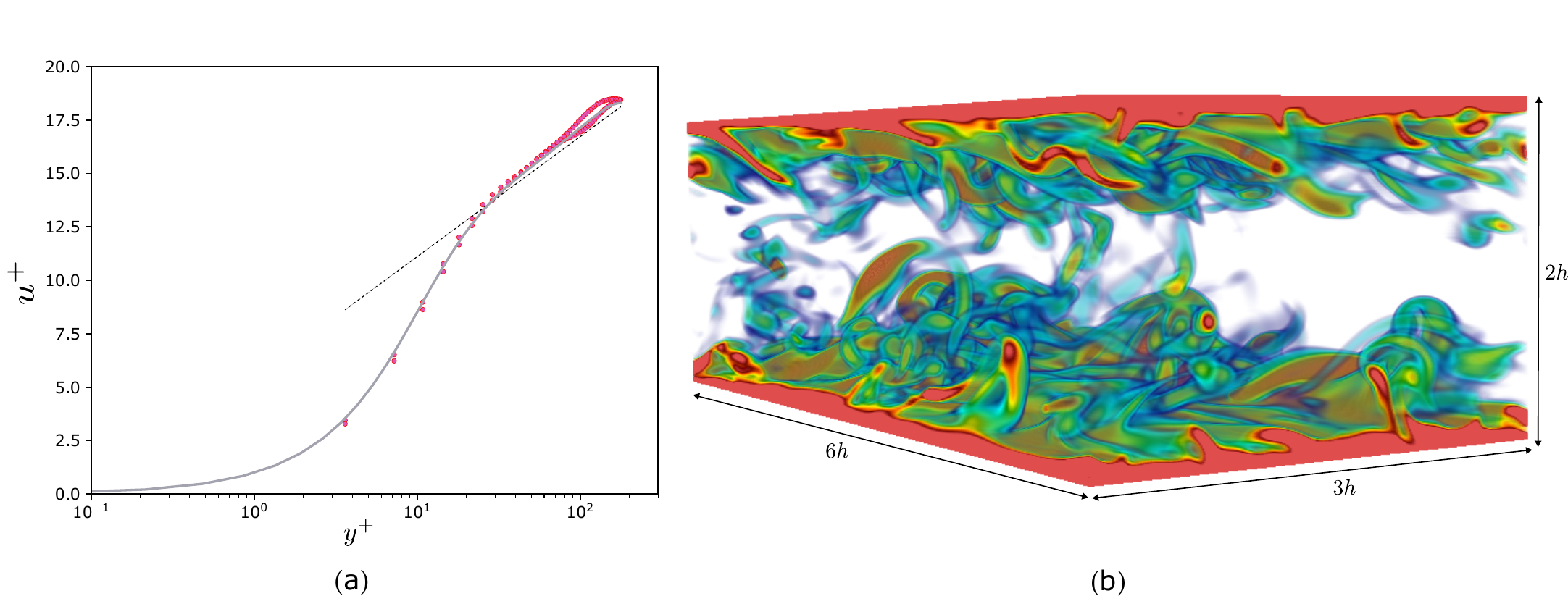}
	\caption{(a) Comparing results taken with $h=50$ using XLB with DNS data of~\cite{Moser:1999:DNS} (solid line), as well as the log-law of wall (dashed line)~\eqref{eq:loglaw}. The KBC collision model \cite{Karlin:2014:gibbs} in XLB was used for this example. }\label{fig:turbChannel} 
\end{figure}

As illustrated in Figure~\ref{fig:turbChannel}(b), we define the channel by two stationary walls that are $2h$ apart. The flow is assumed to be periodic in both the transverse and streamwise directions where the channel extents are assumed to be  $3h$ and $6h$ respectively. The flow is continuously forced in the streamwise direction by $F_x$ defined as,
\begin{equation}
    F_x = \frac{Re_\tau^2  \nu^2}{h^3}
\end{equation}

Figure~\ref{fig:turbChannel} compares XLB results against Direct Numerical Simulation (DNS) results of~\cite{Moser:1999:DNS} at $Re_{\tau} = u_\tau h/ \nu = 180$. Here the distance is measured in wall units and defined as $y^+ =y u_\tau/\nu$ while the averaged velocity (averaged along both periodic directions) is normalized using the wall friction velocity ($u_\tau = \tau_w /\rho$ where $\tau_w$ is the wall shear stress). For reference, we have also included the log-law of wall introduced by von K\'arm\'an as,
\begin{equation}
    u^+ ( y^+) = \kappa^{-1}\log(y^+) + C^+, \qquad \mbox{for}\ y^+ \gg 1
    \label{eq:loglaw}
\end{equation}
in which the von K\'arm\'an constant is $\kappa \approx 0.41$ and $C^+ \approx 5.5$.

The predicted numerical results are again in excellent agreement with reference DNS data even though unlike DNS the scales of motion are not fully resolved down to the dissipation scale in our LES simulation.

%% file: tex/performance.tex
\section{Performance}
\label{sec:performance}

Compared to low-level programming languages like C++, it is not straightforward to incorporate certain performance optimizations in JAX without compromising the readability and flexibility of the codebase. In C++, for example, a user has direct control over many low-level aspects, such as memory layout, explicit SIMD (Single Instruction, Multiple Data) vectorization, and fine-grained control over how data is loaded into cache. These optimizations are required for achieving maximum performance in LBM. However, JAX abstracts away many of these low-level details to simplify the user experience and maintain its high-level API. While this abstraction makes JAX more user-friendly and flexible, especially for conducting complex mathematical operations and automatic differentiation, it can limit the ability to employ specific performance optimizations that are readily accessible to a C++ user. We are actively working to mitigate these limitations in XLB and achieve state-of-the-art performance and aim to add a kernel-based backend to XLB based on Warp~\cite{warp2022}. Below, we present performance characteristics of the current XLB version, which relies on the JAX backend.

To better understand the actual performance capabilities of XLB, especially in the context of these inherent trade-offs and JAX's unique computational model, we have conducted a series of detailed performance evaluations. We discuss the performance characteristics of the XLB library across single GPU, multi-GPU, and distributed systems. The XLB library offers satisfactory performance even on desktop GPUs. Although it does not match the speed of highly optimized LBM codes written in low-level programming languages (see e.g.~\cite{sailfish2014, meneghin2022neon,krause2021openlb,Latt:2021:stlbm}), it is sufficiently fast for most practical use cases. The library is also scalable, allowing for performance scaling when required.

All the performance evaluations are done based on a 3D lid-driven cavity simulation with stationary walls on all sides except the top lid that moves horizontally at a constant velocity. The full-way bounce-back scheme was used to impose $\boldsymbol{u}=(0, 0, 0)$ and the equilibrium boundary condition was employed for the top lid to impose $\boldsymbol{u} = (U, 0, 0)$. The \texttt{D3Q19} lattice and BGK collision model are used, and the simulation is executed without any I/O operations. Unless explicitly stated, all tests are done in single precision. We ignored the first iteration to account for the time required for the JIT compilation of the main loop. The performance is quantified in terms of million lattice updates per second (MLUPS) where a single lattice update corresponds with a full LBM iteration including collision, streaming and boundary condition steps.

The examples folder in the repository contains scripts that can be used to reproduce the performance data presented in this section. It should also be noted that the library's performance may change over time due to ongoing improvements or updates to dependencies like JAX or JAXLIB.\ Users are advised to test the library's performance using the latest version.

\subsection{Single GPU performance}
\label{sec:single_gpu_performance}
\begin{figure}[H]
\centering
\includegraphics[width=\textwidth]{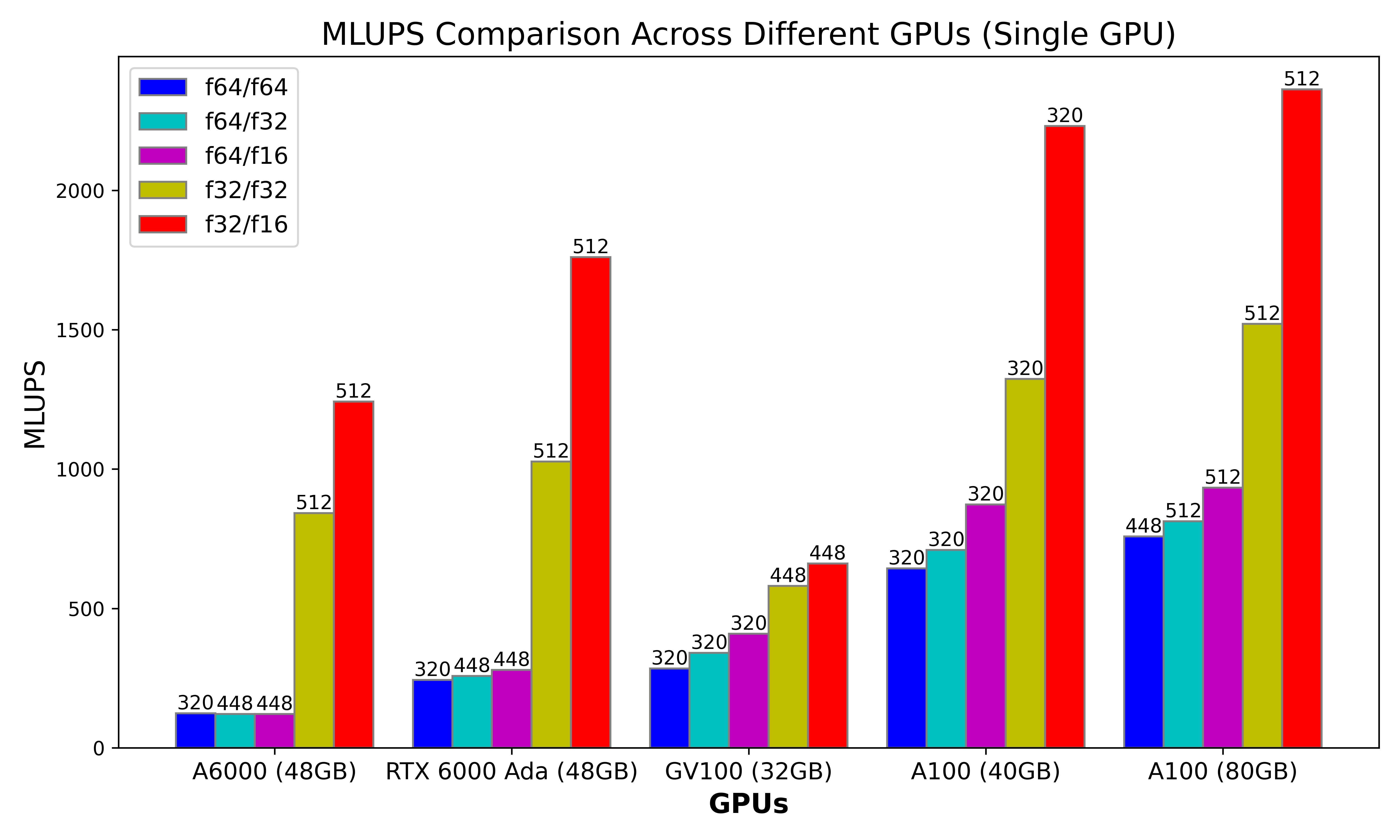}
\caption{Performance evaluation of XLB library on single-GPU systems across various mixed-precision configurations in terms of MLUPS.}\label{fig:MLUPS}
\end{figure}

The performance characteristics of the library on a single-GPU system are shown in Figure~\ref{fig:MLUPS}. The performance metrics have been gathered across various mixed precision configurations and GPU types, including both desktop-grade and server-grade units. Owing to the different memory capacities of these GPUs, we adjusted the domain sizes for the tests. These domain sizes, indicated as labels above each bar in the graph, are sufficiently large to leverage the full performance capabilities of the GPUs (but do not signify the maximum domain size each GPU can handle).

It is evident from the results that desktop-grade GPUs, like NVIDIA's A6000 and RTX 6000 Ada, experience a more substantial decline in double-precision performance. This is attributed to their limited native support for double-precision calculations. Notably, NVIDIA's latest hardware, such as the RTX 6000 Ada Lovelace, excels in low-precision computations, offering more than six times the performance in \texttt{f32/f16} than that of double-precision computations. However, it should be noted that opting for FP16 storage precision may compromise numerical accuracy as discussed in Section~\ref{sec:benchmarks}, which may not be acceptable for certain applications.

\subsection{Multi-GPU scaling}
\label{sec:multi_gpu_scaling}

We highlight the scaling capabilities of the XLB library on a single-node, multi-GPU environment. We conduct these tests on a DGX system equipped with A100 (80GB) GPUs. The test configurations remain consistent with those described in Section~\ref{sec:single_gpu_performance} and the precision is set to be single precision (i.e. \texttt{f32/f32}). Both strong and weak scaling of the system are examined.

\begin{figure}[H]
    \centering
    \includegraphics[width=0.8\textwidth]{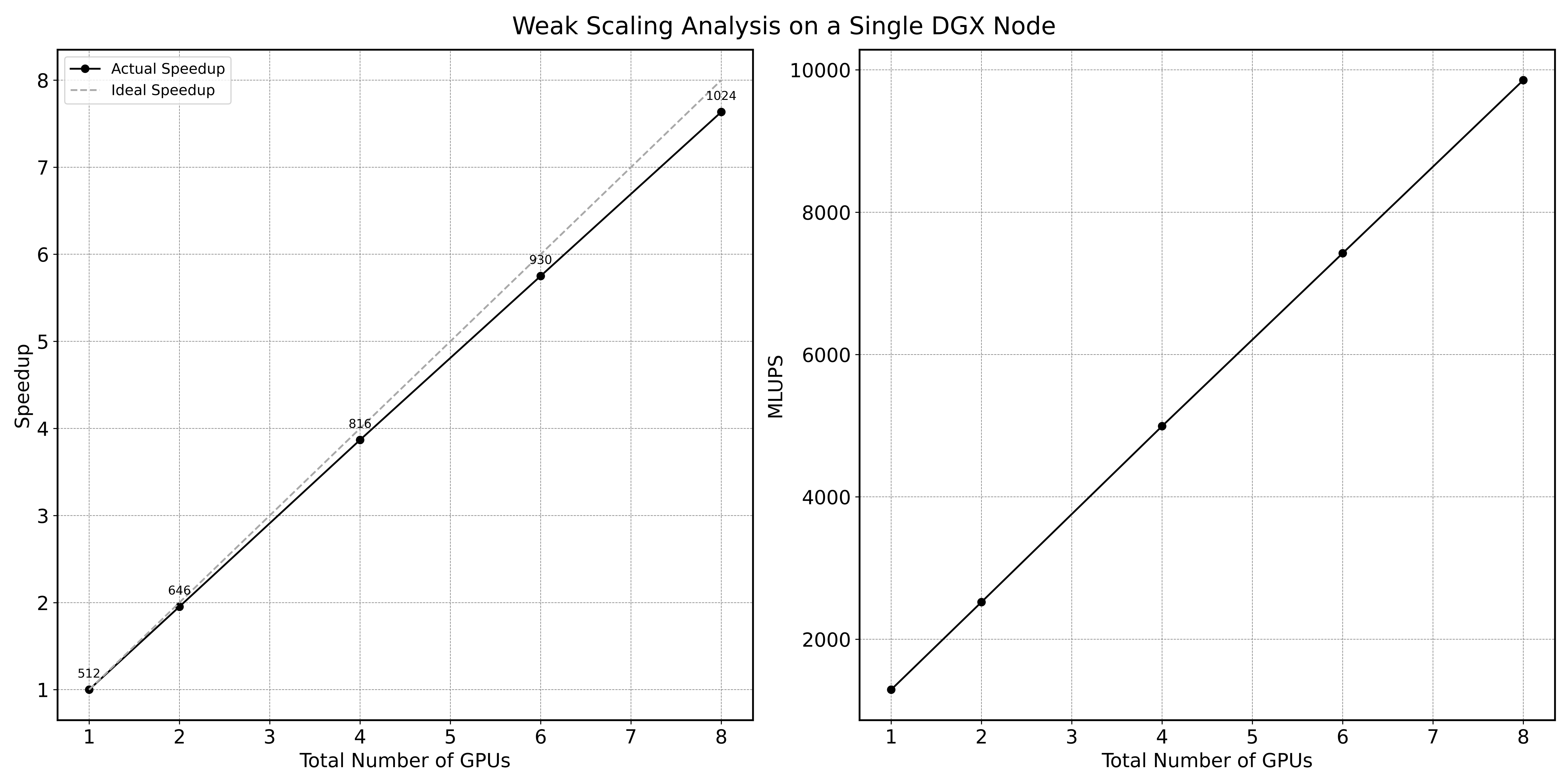}
    \caption{Weak scaling of XLB library on a DGX A100 multi-GPU setup (single precision). Domain sizes are adjusted in proportion to the number of GPUs used (shown with annotations). }\label{fig:weak_scaling}
\end{figure}

Weak scaling is illustrated in Figure~\ref{fig:weak_scaling}. Domain sizes for each data point are indicated through annotations. We note that due to the sharding configuration, domain sizes must be evenly divisible among the GPUs. Therefore, when needed, we round up each domain size to the nearest multiple of the number of GPUs in use. The data reveals excellent scaling efficiency, exceeding 95\% efficiency retention when using 8 GPUs. 

\begin{figure}[H]
    \centering
    \includegraphics[width=0.8\textwidth]{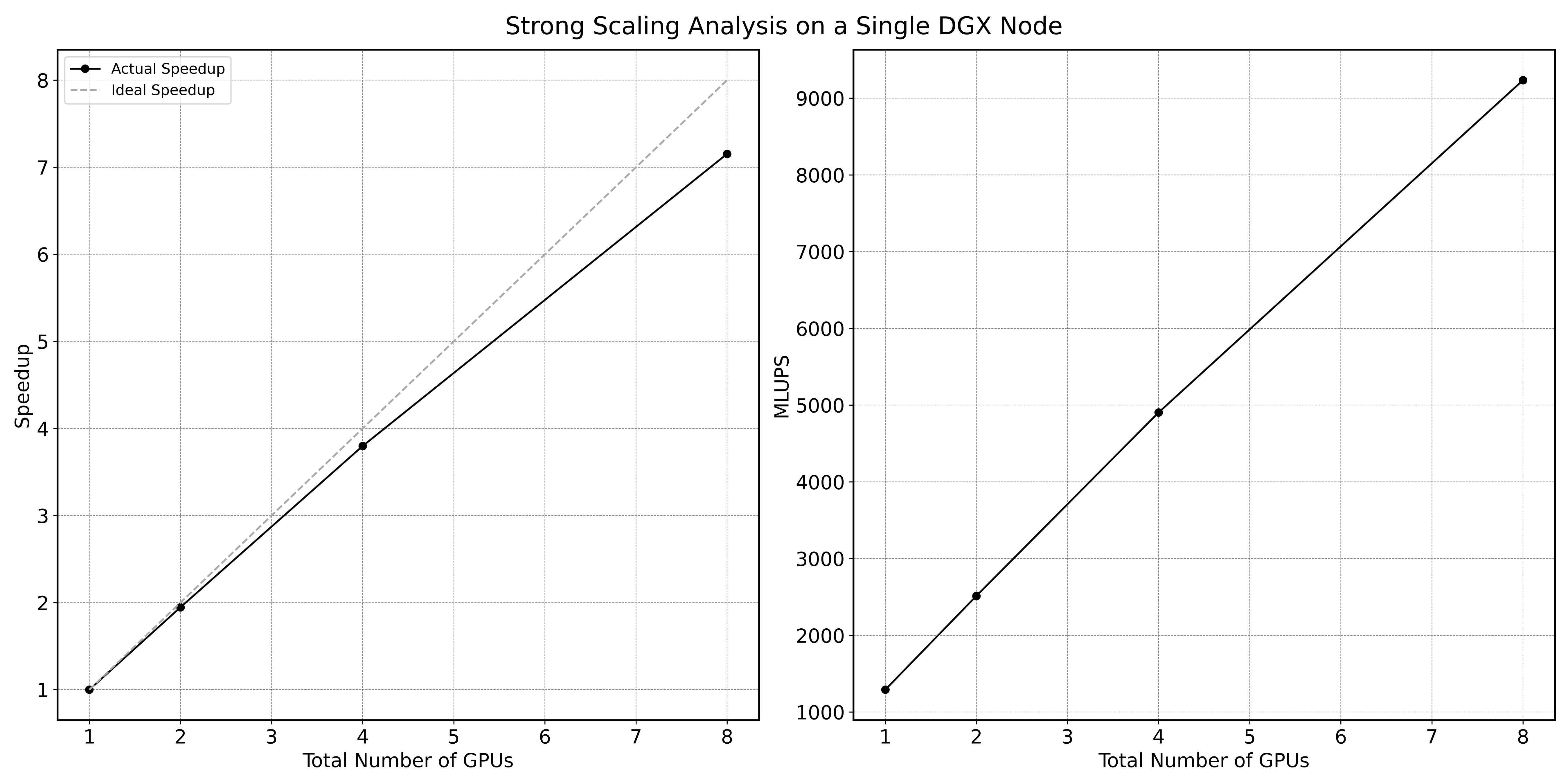}
    \caption{Strong scaling of XLB library with fixed problem size on DGX A100 multi-GPU setup (single precision).}\label{fig:strong_scaling}
\end{figure}

Strong scaling performance is shown in Figure~\ref{fig:strong_scaling}. For this test, the problem size remains fixed at dimensions of $512 \times 512 \times 512$ while the number of GPUs is increased. Similar to the weak scaling tests, the strong scaling results also indicate remarkable efficiency, with close to 90\% efficiency using 8 GPUs. We acknowledge that these multi-GPU scaling results might be accentuated due to the relatively low single-GPU performance of our JAX implementation.

\subsection{Distributed scaling}
\label{sec:multi_node_scaling}

We evaluate the distributed weak scaling abilities of XLB on a cluster of NVIDIA's DGX A100 (80GB) nodes each with 8 GPUs. The scaling test extends up to 64 nodes, equivalent to a total of 512 GPUs, with test configurations maintained consistently as elaborated in preceding sections. MPI can be employed to run each individual JAX process on a separate node. To initiate JAX processes, the \texttt{jax.distributed.initialize} function must be called at the beginning of each process. Aside from these changes, no other modification is necessary to enable running XLB on distributed systems.

\begin{figure}[H]
\centering
\includegraphics[width=1.0\textwidth]{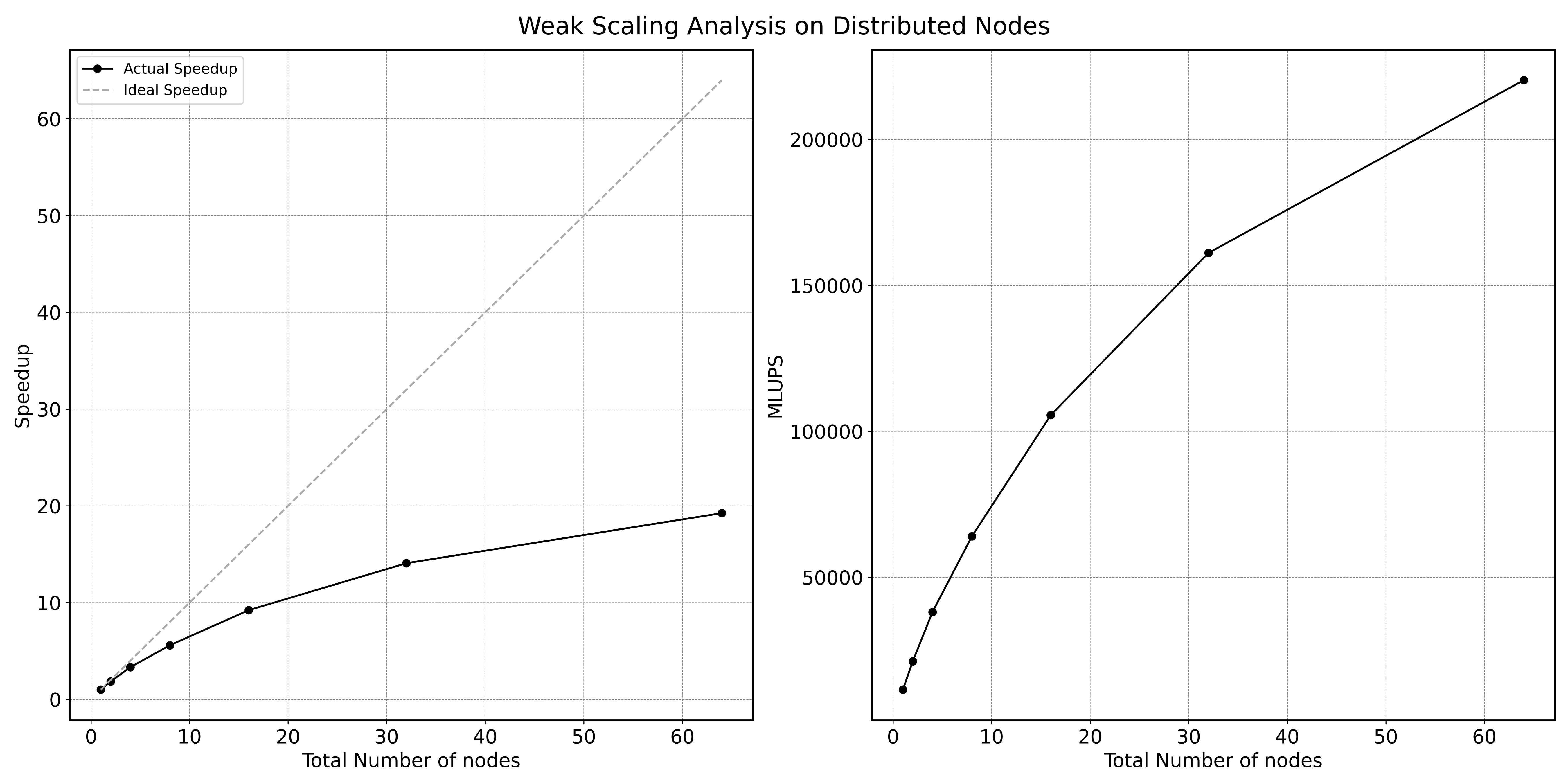}
\caption{Distributed scaling of XLB library on a DGX A100 distributed setup. Domain sizes are adjusted in proportion to the number of GPUs used (shown with annotations).}\label{fig:multi_node_scaling}
\end{figure}

Our results are presented in Figure~\ref{fig:multi_node_scaling}. Remarkably, we observed a relatively good scaling efficiency of 70\% up to eight nodes or 64 GPUs. Beyond this point, however, efficiency began to decline, ultimately reaching approximately 30\% when using 64 nodes, or 512 GPUs. The main cause for this diminishing efficiency is the increase in communication overhead between nodes as the system scales. In the simulation on 64 nodes, the domain size expands to $4096 \times 4096 \times 4096$, having more than 67 billion fluid cells. This exceptionally large simulation size exceeds typical academic and industrial CFD runs, which also underscores the exceedingly high computational and communication demands of such a task. The raw performance numbers of these tests are summarized in Table~\ref{tab:performance_weakscaling}.

\begin{table}[H]
    \centering
    \begin{tabularx}{\textwidth}{XXXXX}
    \toprule
    \textbf{Domain / dimension} & \textbf{Number of nodes} & \textbf{Number of GPUS} & \textbf{MLUPS} \\
    \midrule
    1024 & 1 & 8 & 11448 \\
    1296 & 2 & 16 & 21159 \\
    1632 & 4 & 32 & 38055 \\
    2048 & 8 & 64 & 63987 \\
    2688 & 16 & 128 & 105525 \\
    3328 & 32 & 256 & 161108 \\
    4096 & 64 & 512 & 220332 \\
    \bottomrule
\end{tabularx}
\caption{Weak scaling performance of the XLB library on a cluster of DGX A100 nodes, each with 8 GPUs (single precision). The test extends up to 64 nodes (512 GPUs), with each JAX process running on a separate node using MPI.}\label{tab:performance_weakscaling}
\end{table}

%% file: tex/sciML.tex
\section{Physics-based machine learning}
\label{sec:sci_ml}

XLB's power in scientific machine learning stems from its native differentiability that is provided through integration with JAX, and hence enables a unified computational graph for both forward and backward computations. Linking simulation packages and machine learning models from different libraries often requires complex integration and can lead to performance losses. XLB stands out by allowing for seamless forward and backward propagation across both simulation and neural network components, offering the flexibility to integrate machine learning models at various stages of the simulation workflow, as evidenced in the following demos using the Flax library~\cite{flax2020github}.

In the following, we provide two demos to showcase the capabilities of XLB for investigating novel ideas in physics-based machine learning. These two demos are (1) reducing coarse-grained simulation error with deep learning correctors, and (2) differentiable fluid flow control with deep learning. These examples are intended to highlight the utility and applicability of XLB in scientific contexts, emphasizing its unique features, without being the central focus of our discussion.

\subsection{Enhancing coarse-grained simulations using deep learning correctors}

Drawing on the methodology introduced in~\cite{um2020solver}, this example aims to improve the accuracy of a low-fidelity coarse simulation by applying a neural network corrector. We employ the same example of unsteady flow over a cylinder that was discussed in Section~\ref{sec:flow_over_cylinder}. The corrector is designed to act as a body force that attempts to modify the coarse simulation to resemble the reference solution of the same setup but at a higher resolution (based on an $L^2$ norm).

We define the time-series associated with the evolution of the reference solution at a high-resolution by 
\[
\{\boldsymbol{f}_{h}(\boldsymbol{x}_h,t_0), \, \boldsymbol{f}_{h}(\boldsymbol{x}_h, t_0 + \Delta t_h), \cdots, \boldsymbol{f}_{h}(\boldsymbol{x}_h, t_0 + k\Delta t_h) \} 
\] 
in which $\boldsymbol{f}_{h}(\boldsymbol{x},t)$ corresponds with the distribution function field in all $q$ directions. Here $k$ temporal snapshots are included in the time-series.
The above reference solution may be coarse-grained at a lower resolution by using bi-cubic spatial interpolation. We denote the resulting time-series by, 
\[
    \{ \boldsymbol{f}_{l}(\boldsymbol{x}_l, t_0), \boldsymbol{f}_{l}(\boldsymbol{x}_l, t_0 + \Delta t_l), \cdots, \boldsymbol{f}_{l}(\boldsymbol{x}_l, t_0 + k\Delta t_l) \}
\] 
in which $\boldsymbol{x}_l$ corresponds with a coarse discretization of the domain. Note that the subscript $h$ and $l$ designate high and low resolutions, respectively. To ensure that the simulations remain identical across different scales, we employed acoustic scaling such that $\Delta \boldsymbol{x}_h/\Delta \boldsymbol{x}_l = \Delta t_h/\Delta t_l = r <1$.

Furthermore, we may define $\boldsymbol{g}(\boldsymbol{x}_l,t)$ as the solution of the LBM equations on the coarse grid without external force; a solution that is completely oblivious to the reference solutions $\boldsymbol{f}_{h}(\boldsymbol{x}_h,t)$ or its coarse-grained alternative $\boldsymbol{f}_{l}(\boldsymbol{x}_l,t)$. The resulting time-series may be written similarly as,
\[
    \{ \boldsymbol{g}(\boldsymbol{x}_l, t_0), \boldsymbol{g}(\boldsymbol{x}_l, t_0 + \Delta t_l), \cdots, \boldsymbol{g}(\boldsymbol{x}_l, t_0 + k\Delta t_l) \}
\]
We may parameterize a body force $\mathcal{F}^{nn}$ by a deep neural network and add it to the above coarse simulation which yields the following time-series,
\[
    \{ \boldsymbol{g}^{nn}(\boldsymbol{x}_l, t_0), \boldsymbol{g}^{nn}(\boldsymbol{x}_l, t_0 + \Delta t_l), \cdots, \boldsymbol{g}^{nn}(\boldsymbol{x}_l, t_0 + k\Delta t_l) \}
\] 
The latter is obtained after consecutive LBM steps, each including a sequence of operations that may be denoted collectively by $\mathscr{P}$ which represents collision, applying force, streaming, and applying boundary conditions altogether in order to advance the result one step in time. The deep learning corrector force, $\mathcal{F}^{nn}$, is then developed to minimize the difference between the resulting macroscopic velocity fields obtained based on the down-sampled coarse-grained solution $\{\boldsymbol{f}_{l}(\boldsymbol{x}_l, t)\}$ and the corrected solution $\{ \boldsymbol{g}^{nn}(\boldsymbol{x}_l, t)\}$. This is achieved by minimizing the following loss function $\mathcal{L}$, expressed as:

\begin{align}
    \mathcal{L} = &
        \sum_{m}^{m+s} 
        \left[ \mathscr{U}_x \big(\boldsymbol{g}^{nn}(\boldsymbol{x}_l, t_0 + m\Delta t_l) \big) - \mathscr{U}_x( \boldsymbol{f}_{l}(\boldsymbol{x}_l, t_0 + m\Delta t_l) ) \right]^2 \notag \\[1ex]
        & +
        \sum_{m}^{m+s} 
        \left[ \mathscr{U}_y(\boldsymbol{g}^{nn}(\boldsymbol{x}_l, t_0 + m\Delta t_l)) - \mathscr{U}_y(\boldsymbol{f}_{l}(\boldsymbol{x}_l, t_0 + m\Delta t_l) )\right]^2 
\end{align}
in which $\mathscr{U}_x$ and $\mathscr{U}_y$ represent the operators that extract the $x$ and $y$ components of the velocity field from the distribution functions respectively as per \eqref{eq:velocity}. The initial time step for training is identified by $m$, and $s$ represents the number of time steps that are `unrolled' in the training procedure. Due to the temporal unrolling, the network's output influences subsequent loss calculations. Therefore, backpropagation involves differentiating all the LBM steps combined in the operator $\mathscr{P}$, enabling the network to understand the temporal dynamics of the flow and adjust its output to minimize future losses.

The neural network adjusts the fluid flow dynamics by exerting the corrector force $\mathcal{F}^{nn}(\boldsymbol{x}, t)$. Using the exact-difference method of \cite{kupershtokh2004}, the force term is applied through perturbations in the equilibrium distribution function due to a change in the velocity field $\Delta \boldsymbol{u}^{nn}$ such that $\Delta \boldsymbol{u}^{nn} = \mathcal{F}^{nn} \Delta t / \rho$. More explicitly, in the presence of an external force, this method modifies the post-collision distribution functions, $f_i^\ast(\boldsymbol{x}, t)$, as follows:
\begin{equation}
f_i^\ast(\boldsymbol{x}, t) = \mathscr{C}(f_i(\boldsymbol{x}, t)) + \Delta t \left[ f_{i}^{eq}(\rho, \boldsymbol{u} + \varepsilon \Delta \boldsymbol{u}^{nn}) - f_{i}^{eq}(\rho, \boldsymbol{u})\right],
\end{equation}
in which we have added $\varepsilon=10^{-2}$ as a scaling factor to stabilize the simulation during initial training cycles since the neural network outputs are initialized randomly.

\begin{figure}[hbt!]
    \centering
    \includegraphics[width=\textwidth]{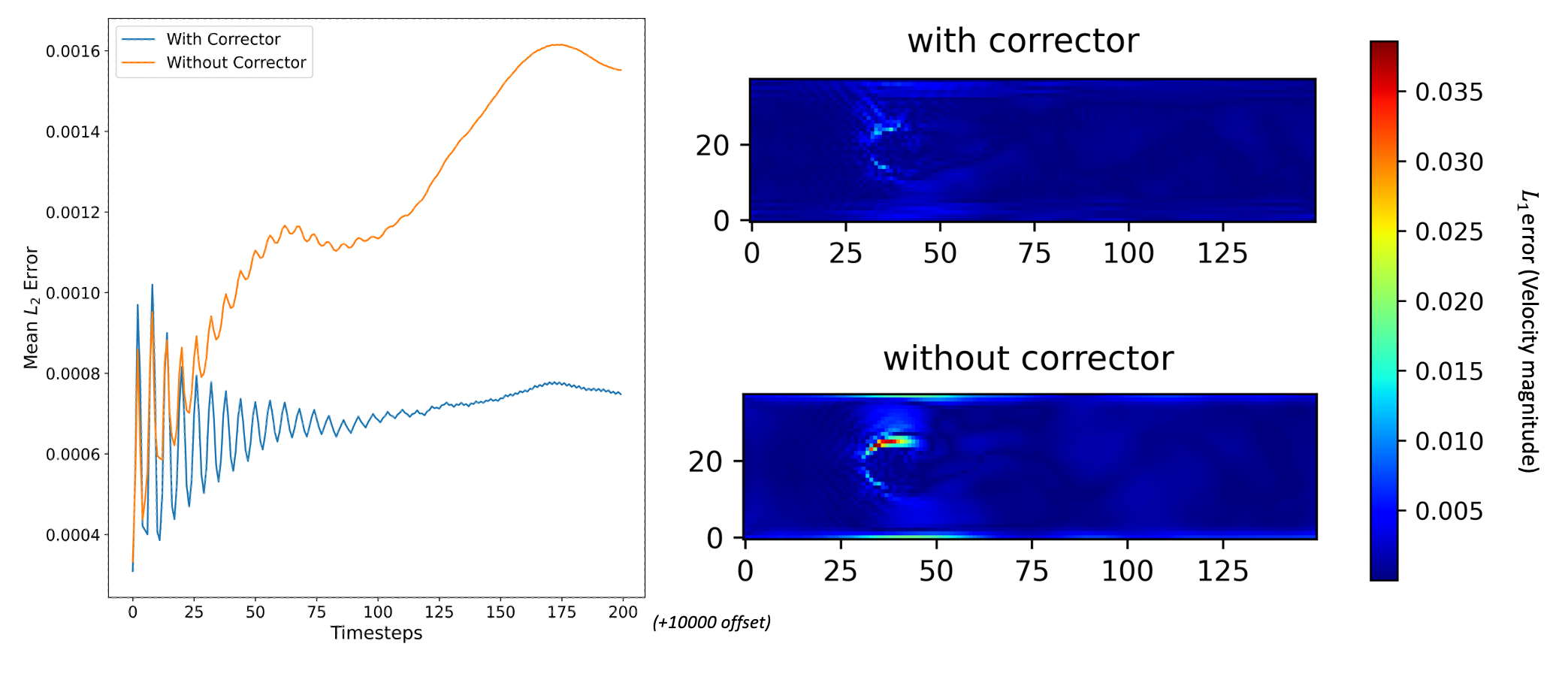}
    \caption{Comparison of simulation error with and without the deep learning corrector over time, demonstrating the corrector's effectiveness in reducing mean error.}
    \label{fig:corrector}
\end{figure}

For training, we adopted a reference simulation at high resolution with $n_x \times n_y = (456\times 120)$, and down-sampled its solution by a factor of $r=6$ to arrive at the coarse-grained reference solution. The deep learning model utilizes a ResNet architecture with four residual blocks, each with a $3\times 3$ kernel size, developed using the Flax library~\cite{flax2020github}. Training cycles begin with the reference solution being used to initialize $\boldsymbol{f}_l$, setting $s=100$ and $m$ to values within $(0, 1, \ldots, 200)$. Gradient checkpointing optimizes memory use during backpropagation. As training cycles are independent, we employ batching over time steps with a batch size of 20. Training starts after  $t_0 = 10,000$ time steps to allow for the development of unsteady vortex shedding in the wake, continuing for $s$ additional time steps. We utilize the Adam optimizer with a learning rate of $10^{-3}$ and train the model for Reynolds numbers $Re= 950, 1000, 1100$, for 50 epochs each, and testing it at $Re=1050$.

Figure~\ref{fig:corrector} illustrates the significant reduction in simulation error achieved by integrating the deep learning corrector, compared to a coarse simulation without the corrector. Both simulations begin with the same reference solution, yet the mean error is notably lower with the inclusion of the corrector as the simulation progresses. 

While promising, these techniques often suffer from generalization issues, and the corrector's performance may degrade when applied to different flow regimes or geometries. It is important to note that parallel-in-time (PIT) methods~\cite{randles2014parallel} offer an alternative approach to enhancing coarse-grained simulations without deep learning. PIT, under certain use cases, may alleviate generalization issues and provide better computational advantages. The work presented here serves as a proof-of-concept for incorporating deep learning models in an LBM simulation, and further research is needed to explore the full potential and limitations of these approaches.

\subsection{Differentiable flow control with deep learning}
\begin{figure}[H]
    \centering
    \includegraphics[width=\textwidth]{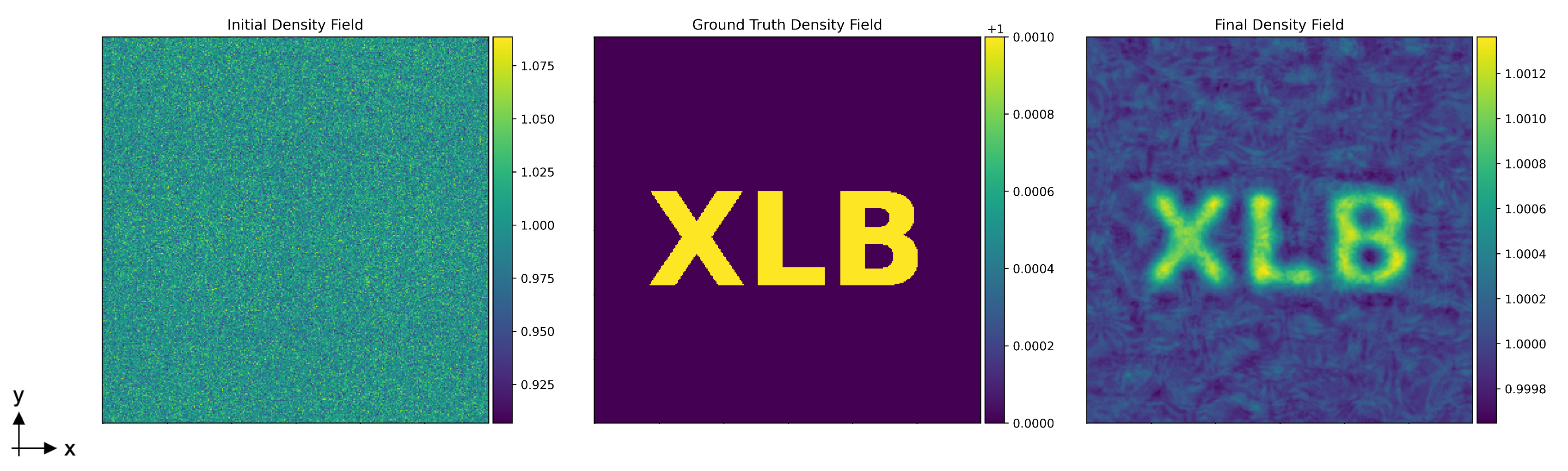}
    \caption{Depiction of the initial density field, the targeted ``XLB'' density pattern as the ground truth, and the resultant density field after 200 iterations of the optimization process.}\label{fig:XLB}
\end{figure}

In this second demo, we aim to solve an inverse flow control problem. In particular, we seek to determine an initial condition for the evolving nonlinear dynamics of the fluid flow that leads to the creation of a pressure field resembling the visualization of the word ``XLB'' (see Figure~\ref{fig:XLB}) after $k$ LBM time steps under periodic boundary conditions. This problem involves optimizing the initial conditions such that the density field evolves to display the pattern at a pre-determined time step.

The loss function for this optimization is given by:
\begin{equation}
    \mathcal{L} = \left(\sum_{i=1}^{q}f^k_{i} - \rho_{\text{xlb}}\right)^2
    \label{eq:loss}
\end{equation}
where $k$ is the target time step for the ``XLB'' pattern to become visible, $f^k_{i}$ is the density distribution function along the $i$th lattice direction at time step $k$, and $\rho_{\text{xlb}}$ represents a prescribed density field where the word ``XLB'' is encoded by having a standard density of 1.0 with an increase to 1.001 at the positions forming the letters.

In our experiment, we arbitrarily set $k=200$. The simulation is based on a 2D square with periodic boundaries on all sides and a resolution of $300\times300$. We chose a simple feedforward multi-layer perceptron consisting of three layers with 32, 64, and 32 neurons, respectively, as a baseline for this proof-of-concept experiment. The neural network takes in the initial density and perturbs it. The Adam optimizer, with a learning rate of $10^{-3}$, is utilized to minimize the loss function across 300 epochs of training.

\begin{figure}[H]
    \centering
    \includegraphics[width=\textwidth]{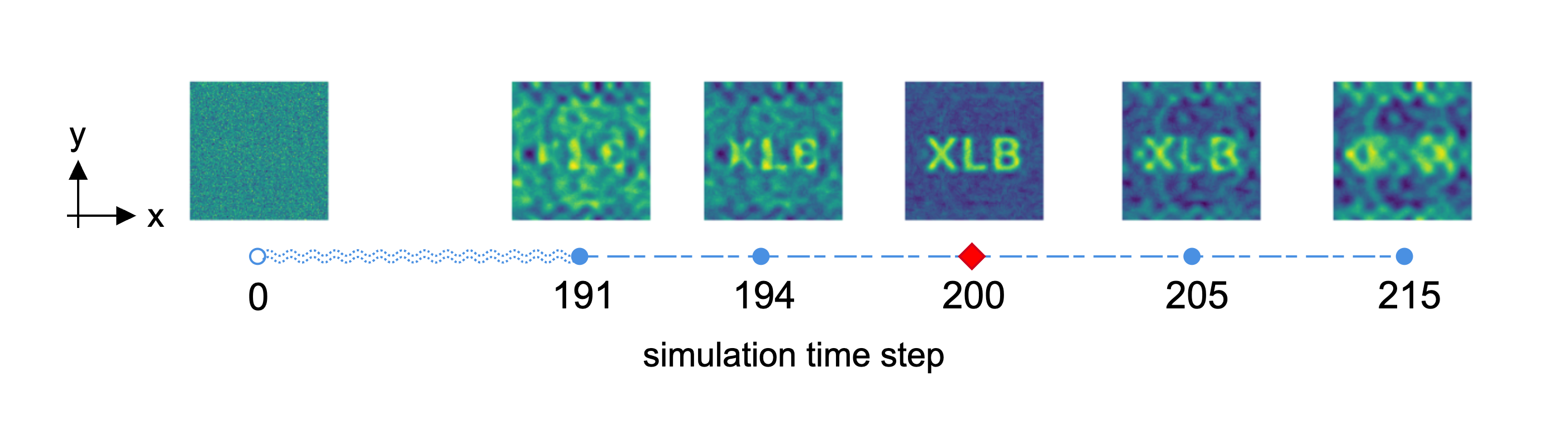}
    \caption{Temporal progression of the fluid's density field, depicting the initial state and the formation of the ``XLB'' pattern at the designated time step $k=200$, with select intermediate states.}\label{fig:XLB_evolution}
\end{figure}

The results obtained after solving the above optimization problem are illustrated in Figures~\ref{fig:XLB} and \ref{fig:XLB_evolution}. Aside from the qualitatively satisfactory outcome of this demo, this experiment demonstrates as a proof-of-concept how automatic-differentiation embedded in the design of XLB library can be easily used to solve more complex inverse problems in CFD. While the initial density distribution obtained here may be considered as a noise field and hence not physically viable due to small-scale perturbations to $\rho$, the loss function in Eq.~\ref{eq:loss} could be refined with physically-informed constraints to identify better initial conditions that yield more realistic flow patterns.

%% file: tex/conclusion.tex
\section{Conclusion and future work}
\label{sec:conclusion}

This paper presented XLB library, a distributed multi-GPU differentiable LBM library based on JAX, tailored for large-scale fluid simulations and physics-based machine learning. We have verified the accuracy and reliability of XLB through a series of benchmarks. Additionally, we evaluated the parallel performance of XLB and demonstrated its efficient scaling to run on many GPU devices. We also highlighted some applications of XLB for integrating machine learning in fluid mechanics, with examples on reducing simulation errors through deep learning correctors and controlling fluid flow via deep learning techniques.

As we look to the future, the XLB library is set to evolve significantly. The roadmap for development is rich with ambitious features, including but not limited to adjoint-based optimization, data assimilation, machine learning-based simulation acceleration, different backends and programming models (e.g.\ based on Warp~\cite{warp2022} and Neon~\cite{meneghin2022neon}) to achieve state-of-the-art performance, and to enable grid refinement (see~\cite{Mahmoud:2024:OGI}) and out-of-core computing strategies in order to handle even larger datasets and simulations. These enhancements, among many others, are currently in various stages of planning and development. 

This document marks the initial introduction of the XLB library, with a strong emphasis on community-driven contributions for its ongoing enhancement. The continuous improvement and success of this open-source project hinge on active participation from the scientific community. Therefore, we warmly invite researchers and developers to engage with the XLB library, contribute their insights and enhancements, and stay connected with its progress by following the project's repository.

%% file: tex/ack.tex
\section*{Acknowledgments}
\noindent The authors wish to express their gratitude to the Nvidia JAX team for their invaluable support and for generously providing the computational resources essential for performance testing of XLB on distributed GPU clusters. We would like to especially acknowledge and appreciate several helpful tips and comments by Fr\'ed\'eric Bastien at NVIDIA who also oversaw the distributed scaling results presented in Section~\ref{sec:performance}, and Arun Raman for doing the scaling runs. We would also like to acknowledge Massimiliano Meneghin, Ahmed Mahmoud, Olli Lupton, and Santosh Bhavani for their valuable review of the final manuscript.